\documentclass[aps,amsmath,twocolumn,amssymb,titlepage,10pt,superscriptaddress]{revtex4-1}
\usepackage[T1]{fontenc}
\usepackage[utf8]{inputenc}
\usepackage{amsmath}
\usepackage{braket}
\usepackage{amsfonts}
\usepackage{comment}
\usepackage{graphicx}
\usepackage{hyperref}
\usepackage[capitalize]{cleveref}
\usepackage{amssymb}
\usepackage{amsmath}
\usepackage{mathtools}
\usepackage{todonotes}

\newcommand{\ketbra}[2]{\ket{#1}\!\bra{#2}}

\begin{document}
\title{Thermodynamics of the spin-1/2 Heisenberg antiferromagnet on the star lattice}
\author{Adrien Reingruber}
\affiliation{Institute for Theoretical Solid State Physics, RWTH Aachen University, JARA Fundamentals of Future Information Technology, and \\ JARA Center for Simulation and Data Science, 52056 Aachen, Germany}
\affiliation{Institut für Theoretische Physik und Astrophysik, Universität Würzburg, Am Hubland, D-97074 Würzburg, Germany}
\author{Nils Caci}
\affiliation{Institute for Theoretical Solid State Physics, RWTH Aachen University, JARA Fundamentals of Future Information Technology, and \\ JARA Center for Simulation and Data Science, 52056 Aachen, Germany}
\author{Stefan Wessel}
\affiliation{Institute for Theoretical Solid State Physics, RWTH Aachen University, JARA Fundamentals of Future Information Technology, and \\ JARA Center for Simulation and Data Science, 52056 Aachen, Germany}
\author{Johannes Richter}
\affiliation{Institut für Physik, Universität Magdeburg, P.O. Box 4120, 39016 Magdeburg,Germany}
\affiliation{Max-Planck-Institut für Physik Komplexer Systeme, Nöthnitzer Str. 38, 011087 Dresden, Germany}
\begin{abstract}
Using a combination of quantum Monte Carlo simulations in adapted cluster bases, the finite temperature Lanczos method, and an effective Hamiltonian approach, we explore the thermodynamic properties of the spin-1/2 Heisenberg antiferromagnet on the star lattice. We consider various  parameter regimes on this strongly frustrated Archimedean lattice, including the case of homogeneous couplings as well as the distinct parameter regimes of dominant vs. weak dimer coupling. For the latter case, we  explore the quantum phase diagram in the presence of inhomogeneous trimer couplings, preserving inversion symmetry. We compare the efficiency of different cluster decoupling schemes for the quantum Monte Carlo simulations in terms of the sign problem, contrast the thermodynamic properties to those of other strongly frustrated quantum magnets, such as the kagome lattice model, and comment on previous results from tensor-network calculations regarding a valence bond crystal phase in the regime of weak dimer coupling. Finally, we  relate our results to recently reported experimental findings 
on a Cu-based quantum magnetic spin-1/2 compound with an underlying star lattice structure.

\end{abstract}
\maketitle

\section{Introduction}\label{Sec:Introduction}
Among  the eleven Archimedean lattices, i.e., periodic tessellations of the plane by regular polygons such that all edge lengths are equal and every vertex looks alike, several cases are particularly prominent in the field of quantum
magnetism~\cite{Richter2004,Balents2010,farnell2014quantum,Starykh2015,Savary2016}. These include the well-known examples of the square, triangular, honeycomb and kagome lattice. Various magnetic compounds realize the topology underlying these lattices in the form of exchange paths between the magnetic ions. For the cases of the triangular and the kagome lattice in particular, antiferromagnetic exchange interactions induce a strong degree of geometric frustration, resulting from the triangles contained in these lattices. While the spin-1/2 Heisenberg model on the triangular lattice has been concluded to nevertheless still exhibit a magnetically ordered ground state (see Ref.~\cite{Richter2004} and references herein), the stronger geometric frustration along with a lower coordination number of $z=4$ leads to the complete breakdown of classical long-range order in the spin-1/2 Heisenberg model on the kagome lattice, offering a promising candidate for a quantum spin liquid ground state  instead~\cite{Savary2016,Yan2011,Depenbrock2012,Jiang2012,He2017}. 

\begin{figure}[t]
    \centering
    \includegraphics[width=0.4\textwidth]{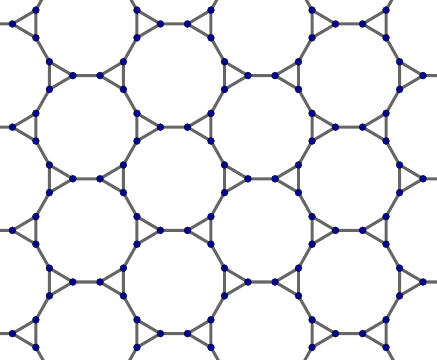}
    \caption{Illustration of the star lattice, also known as the truncated hexagonal, triangle-decorated honeycomb, or T9 lattice, 
    specified by the Gr\"unbaum-Shephard~\cite{Gruenbaum1977} symbol $(3, 12^2)$ in 
terms of the set of polygons which surround each lattice site (one triangle and two dodecagons in this case).}
    \label{Fig:lattice}
\end{figure}

It is worthwhile to note that the star lattice shown in Fig.~\ref{Fig:lattice}  has in fact the lowest coordination number, $z=3$, among all  the Archimedean lattices that contain triangles.
In contrast to other Archimedean lattices, material realizations of the star lattice structure in magnetic compounds are however much less 
abundant. 
Indeed, only two explicit realizations of this 
 lattice structure 
in layered magnetic materials have been  reported~\cite{zheng2007star,Sorolla2020}. Of particular interest here is the compound [(CH${}_3$)${}_2$(NH${}_2$)]${}_3$[Cu${}_3$($\mu_3$-OH)($\mu_3$-SO${}_4$)($\mu_3$-SO${}_4$)${}_3$]·0.24H${}_2$O, which realizes a Cu-based spin-1/2 system with antiferromagnetic interactions and no magnetic order down to 2 K~\cite{Sorolla2020}. We comment further on this material in the conclusions. 

Theoretically, quantum spin models on the star lattice have been investigated in several previous studies with respect to the ground state properties. Below, we summarize previous studies of the SU(2)-symmetric Heisenberg model on the star lattice  (other works  focus on quantum spin liquid phases on the star lattice in models with strongly anisotropic, Kitaev-model interactions~\cite{Yao2007, Hickey2021}).
In Refs.~\cite{Richter2004,richter2004starlattice},  the ground state and spectral properties
of the spin-1/2 Heisenberg model on
the  star lattice have been reported based on exact diagonalization (ED) on finite
lattices with up to
42 spins. These studies, which considered the exchange couplings along the bonds of the star lattice of equal strength $J$, 
reported a gapped paramagnetic ground state with a sizeable spin gap (singlet-triplet gap) of
$\Delta=0.3809J$, where this value is an estimate obtained by finite-size
extrapolation of ED data for lattices of $N=18,24,30$ and $36$ sites. 
Whereas the triangles in the kagome lattice are corner sharing, they are coupled by dimer bonds on the star lattice, 
cf. Fig.~\ref{Fig:lattice}. Thus, the latter is an Archimedean lattice with two
non-equivalent nearest-neighbor bonds,  namely triangular  ($J_t$) and dimer ($J_d$)
couplings.  
The spin-spin correlations along these dimer bonds were found to be more than three times larger than on the triangular bonds (in the following, we will also denote these triangles as trimers). In this regime, the ground state is thus strongly dimerized on the dimer (i.e., intertrimer) bonds. Besides the kagome lattice, the star lattice thus provides another example of an Archimedean lattice for which the interplay of geometric frustration and quantum fluctuations prevents ground-state magnetic order to emerge. 

More recently, the case that the strengths of the  intra-trimer ($J_t$) and intertrimer ($J_d$) couplings of the spin-1/2 Heisenberg model on the star lattice take on different values has been considered based on various 
methods~\cite{Misguich2007, Yang2010, Richter2017,Jahromi2018,Ran2018}. 
According to these studies, the aforementioned dimerized phase is stable  up to a ratio of $J_t/J_d\approx 1.1$. The infinite projected entangled pair states (iPEPS) results from Ref.~\cite{Jahromi2018} indicate that for larger values of $J_t$,  the ground state is a valence bond solid (VBS) with a six-site unit cell, spontaneously breaking the C${}_3$ symmetry around each of the trimers.

Here, we extend these previous studies by examining the finite-temperature thermodynamics of the spin-1/2 Heisenberg model on the star lattice over a wide range of the coupling ratio $J_t/J_d$. Furthermore, we also examine this model in the thus-far unexplored regime where the trimer couplings are not all equal.  
In order to access the thermodynamic properties of this system, we use a combination of effective low-energy models derived from perturbation theory in the regime of weak intertrimer coupling, ED, the finite temperature Lanczos method 
(FTLM)~\cite{JaP:PRB94,JaP:AP00,ScW:EPJB10}, as well as  quantum Monte Carlo (QMC) simulations using the stochastic series expansion (SSE) framework
~\cite{Sandvik1991,Sandvik1992,
Sandvik1999, Syljuasen2002, Alet2005}. 
In fact, a further motivation for this study was to examine different recently-developed cluster-based SSE QMC algorithms applied to this frustrated quantum magnet~\cite{Nakamura1998, Honecker2016, Alet2016, Ng2017, Stapmanns2018, Weber2022B, Weber2022A}. More specifically, we examine different SSE QMC algorithms formulated in the single-spin, spin-dimer and spin-trimer computational basis, respectively, and compare 
the severeness of the 
QMC sign problem~\cite{Henelius2000, Troyer2005}  for  these different algorithms. 

The remainder of this paper is organized as follows: In the following Sec.~\ref{Sec:Model}, we introduce the  Hamiltonian of the spin-1/2 Heisenberg model on the star lattice and  review the different QMC methods considered here. We also summarize shortly the FTLM approach as applied to such two-dimensional quantum spin systems.  In Sec.~\ref{Sec:Sign}, we discuss 
the sign problem of the QMC method applied to this model for the various computational bases. Our numerical results for the physical properties of the star lattice model are presented in Sec.~\ref{Sec:Results}, along with perturbation theory results for the low-energy effective models in the weak intertrimer coupling regime. Final conclusions are given in Sec.~\ref{Sec:Conclusions}, along with remarks on the compound [(CH${}_3$)${}_2$(NH${}_2$)]${}_3$[Cu${}_3$($\mu_3$-OH)($\mu_3$-SO${}_4$)($\mu_3$-SO${}_4$)${}_3$]·0.24H${}_2$O.

\section{Model and Methods}\label{Sec:Model}

\begin{figure}[t]
    \centering
    \includegraphics[width=0.4\textwidth]{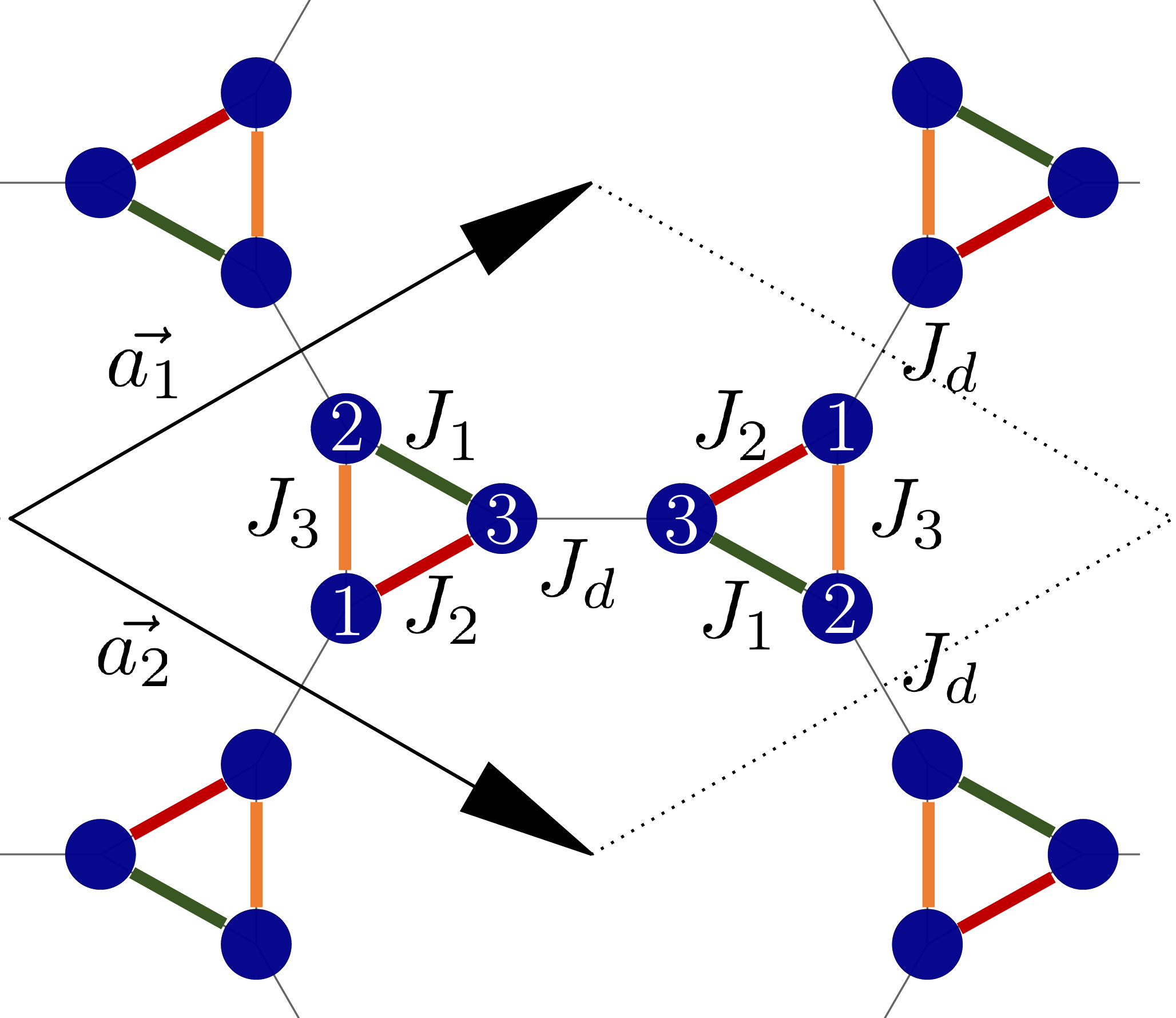}%
    \caption{Unit cell of the star lattice.  The labeling of the sites and the trimer couplings within the  two trimers are indicated as well as the lattice vectors and the dimer couplings.}
    \label{Fig:unitcell}
\end{figure}

In the following we consider the spin-1/2 Heisenberg model with antiferromagnetic exchange interactions on the star lattice, shown in Fig.~\ref{Fig:lattice}. This Archimedean lattice can be considered as a hexagonal lattice of coupled triangles, also referred to as trimers. We denote the  left (right) trimer in each unit cell by $\Delta$ ($\Delta'$), where a given unit cell is specified by its center position $\Vec{r}$. The two lattice vectors that connect neighboring unit cells are denoted by $\Vec{a}_1$ and $\Vec{a}_2$, respectively.  Each unit cell contains two trimers, i.e., a total of six spins, as illustrated in Fig.~\ref{Fig:unitcell}.  In the following, 
we  report QMC results for finite systems with $L\times L$ such unit cells 
(i.e., $N=6L^2$ lattice sites) and with periodic boundary conditions in both lattice directions. 
For the FTLM approach we use a set of finite lattices with
$N=24,30,36,42$ including also finite lattices with  
parallelogram shapes, see Refs.~\cite{Richter2004,richter2004starlattice}.     

The three spins within each trimer are coupled by the trimer interactions $J_1$, $J_2$, and $J_3$. These are labeled such that the spin with index $i$ in a given trimer is located  opposite to the bond with coupling strength $J_i$, cf. Fig.~\ref{Fig:unitcell}. While in general a total of six different trimer couplings are contained in each unit cell, here we consider the inversion symmetric case, cf. Fig.~\ref{Fig:unitcell}, i.e., all couplings along parallel trimer bonds are of equal strength. The trimer Hamiltonian is thus given by
\begin{equation}
H_{\Delta}^{\Vec{r}} = J_1\:\mathbf{S}_{2,\Delta}^{\Vec{r}}\cdot \mathbf{S}_{3,\Delta}^{\Vec{r}}
+ J_2\:\mathbf{S}_{1,\Delta}^{\Vec{r}}\cdot \mathbf{S}_{3,\Delta} ^{\Vec{r}} + J_3\:\mathbf{S}_{1,\Delta}^{\Vec{r}}\cdot \mathbf{S}_{2,\Delta} ^{\Vec{r}},
\end{equation}\label{eq:OneTrimerH}
and the same expression applies for $H_{\Delta'}^{\Vec{r}}$, after replacing $\Delta$ by $\Delta'$. Here, $\mathbf{S}_{i,\Delta}^{\Vec{r}}$ denotes the spin $i$ in trimer $\Delta$ of the unit cell with coordinates $\Vec{r}$.
Spins in neighboring trimers are coupled by the dimer interactions $J_d$, also referred to as intertrimer interactions.   These dimer bonds of the star lattice assign each spin to a unique partner to form a two-site spin dimer (cf. Fig.~\ref{Fig:lattice}). For each unit cell, one dimer bond is contained within the unit cell. The two other dimer bonds connect trimers in neighboring unit cells. More specifically, the  full system's Hamiltonian is  given by
\begin{eqnarray}
H =  \sum_{\Vec{r}}&& \!\! H_{\Delta}^{\Vec{r}} + H_{\Delta '}^{\Vec{r}} 
+ J_d\: \mathbf{S}_{3,\Delta} ^{\Vec{r}}\cdot \mathbf{S}_{3,\Delta'} ^{\Vec{r}}  \nonumber \\ 
&& +J_d\: \mathbf{S}_{1,\Delta'} ^{\Vec{r}}\cdot \mathbf{S}_{1,\Delta} ^{\Vec{r} + \Vec{a_1}} +  J_d \:\mathbf{S}_{2,\Delta'} ^{\Vec{r}}\cdot \mathbf{S}_{2,\Delta} ^{\Vec{r} + \Vec{a_2}}.\label{Eq:ham}
\end{eqnarray}

In order to extract  thermodynamic properties of this model on small finite systems, one can use ED and to access larger system sizes, the FTLM approach. 
Here, we provide only a brief illustration of the basic elements of the FTLM. 
For a more detailed description of the FTLM we refer the interested reader to the reviews~\cite{PrB:SSSSS13,PRE:COR17,ScW:EPJB10} and to recent FTLM papers on the kagome~\cite{kago42} and the square-kagome Heisenberg antiferromagnets \cite{squago_TD_2022}.  

Within the FTLM approach,
the sum over a set of complete orthonormal basis states  in the partition function $Z$ is replaced by
a substantially smaller sum over $R$ random vectors
\begin{eqnarray}
\label{Z}
Z\approx\sum_{\gamma=1}^{\Gamma}
\frac{\dim({\cal{H}}(\gamma))}{R}
\sum_{\nu=1}^{R}\sum_{n=1}^{N_{\rm L}}
{\rm e}^{-\frac{\epsilon_n^{(\nu)}}{T}}
\left\vert\langle n(\nu)\vert\nu\rangle\right\vert^2,
\end{eqnarray}
where $\vert\nu\rangle$ labels the random vectors for each orthogonal subspace ${\cal{H}}(\gamma)$ 
of the Hilbert space with $\gamma$ labeling the respective symmetry.
In Eq.~(\ref{Z}), we approximate the 
exponential of the Hamiltonian by its spectral
representation in a Krylov space
spanned by the
$N_L$ Lanczos vectors starting from the respective random vector
$\ket{\nu}$, where
$\ket{n(\nu)}$ is the $n$-th eigenvector of $H$ in
this Krylov space
with the energy $\epsilon_n^{(\nu)}$.
To perform the symmetry-decomposed numerical Lanczos calculations we
used J\"org Schulenburg's publicly available  package {\it spinpack}
\cite{spin:259,richter2010spin}. A detailed discussion of 
the accuracy of the FTLM can  be found 
in Refs.~\cite{kago42} and \cite{Accuracy2020}. 

Another powerful  approach for the theoretical study of quantum magnetism is provided by unbiased QMC simulations based, e.g.,  on the SSE approach~\cite{Sandvik1991, Sandvik1992, Sandvik1999, Syljuasen2002, Alet2005}. While this method yields highly accurate data for the thermodynamic properties of many quantum magnetic systems, it generally suffers from a 
severe sign problem in the presence of geometric frustration~\cite{Henelius2000, Troyer2005}. Over the recent years, several approaches have been put forward in order to reduce, or in specific cases even remove the 
QMC sign problem in frustrated quantum magnets. One particular  approach is based on reformulations of the QMC sampling scheme by changing from the standard computational basis of local spin  to appropriate cluster bases, such as formed by spin-dimers or spin-trimers~\cite{Nakamura1998, Honecker2016, Alet2016, Ng2017, Stapmanns2018, Weber2022B, Weber2022A}. More specifically, in the local spin basis, referred to as the site basis in the following,
the $S^z$-component of each spin  is diagonal, with
\begin{equation}
S_{j}^z \ket{m_j} = m_j \ket{m_j},
\end{equation}
where the two possible states of the $j$-th spin are denoted by $\ket{\uparrow_j}$ and $\ket{\downarrow_j}$ for $m_j=1/2$ and $m_j=-1/2$, respectively. For this standard computational basis, the formulation of the SSE QMC algorithm has been presented in several works, such as the review in Ref.~\cite{Sandvik2010}.

In the dimer basis, the total spin $\mathbf{S}_d=\mathbf{S}_{j} +\mathbf{S}_{k}$ of the two spins connected by an $J_d$-dimer bond $d$ is considered (note that the two spins belong to different trimers, depending on which specific $J_d$-dimer bond $d$ is considered, cf. Fig.~\ref{Fig:unitcell}). 
The local Hilbert space for the dimer $d$ is then spanned by  four states, consisting of a singlet state
\begin{equation}
\ket{0,0} = \frac{1}{\sqrt{2}}(\ket{\uparrow_j \downarrow_k} - \ket{ \downarrow_j \uparrow_k}),
\end{equation}
and the three  triplet states
\begin{eqnarray}\label{eq:Dimer_Triplet}
&&\ket{1,1} = \ket{\uparrow_j \uparrow_k},\nonumber\\
&&\ket{1,0} = \frac{1}{\sqrt{2}}(\ket{\uparrow_j \downarrow_k}+\ket{\downarrow_j \uparrow_k}),\\
&&\ket{1,-1} = \ket{\downarrow_j \downarrow_k},\nonumber
\end{eqnarray}
where the first (second) quantum number specifies the  eigenvalue of $\mathbf{S}_d^2$ ($S^z_d$) of the total dimer spin.  For an isolated $J_d$ dimer, these four spins form the eigenstates of the Heisenberg Hamiltonian, with the singlet being the ground state. We refer to Ref.~\cite{Honecker2016} for details of the QMC SSE algorithm in the dimer basis. 

Finally, in the  trimer basis  the total spin of one trimer $\mathbf{S}_{\Delta} =  \mathbf{S}_{1,\Delta}+\mathbf{S}_{2,\Delta}+\mathbf{S}_{3,\Delta}$ is considered (here and in the following, we suppress the position vector whenever a specific trimer is considered). We denote the quantum numbers of $\mathbf{S}_{\Delta}^2$ and $S^z_\Delta$ by $l_{\Delta}$, 
and $m_{\Delta}$, respectively. As a further operator to distinguish the eight states of a trimer we use, following Ref.~\cite{Weber2022B}, the operator $\mathbf{S}_{1,2,\Delta} = \mathbf{S}_{1,\Delta} + \mathbf{S}_{2,\Delta}$, and denote by $l_{1,2}$ the quantum number of $\mathbf{S}_{1,2,\Delta}^2$. The basis states for the local Hilbert space for trimer $\Delta$
are therefore denoted by $\ket{l_{1,2},l_{\Delta},m_{\Delta}}$, and  read explicitly
 \begin{equation}\label{eq:Dubletts_D0}
  \begin{split}
   & \ket{0,\tfrac{1}{2},\tfrac{1}{2}} = \frac{1}{\sqrt{2}}(\ket{\uparrow_1 \downarrow_2} - \ket{ \downarrow_1 \uparrow_2})\otimes \ket{\uparrow_3}, \\
  & \ket{0,\tfrac{1}{2},-\tfrac{1}{2}} = \frac{1}{\sqrt{2}}(\ket{\uparrow_1 \downarrow_2} - \ket{ \downarrow_1 \uparrow_2})\otimes \ket{\downarrow_3},
   \end{split}
\end{equation}
and
 \begin{equation}\label{eq:Dubletts_D1}
  \begin{split}
  & \ket{1,\tfrac{1}{2},\tfrac{1}{2}} = \frac{1}{\sqrt{6}}(\ket{\uparrow_1 \downarrow_2 \uparrow_3} + \ket{ \downarrow_1 \uparrow_2 \uparrow_3} - 2\ket{\uparrow_1 \uparrow_2 \downarrow_3 }),\\  
  & \ket{1,\tfrac{1}{2},-\tfrac{1}{2}} = \frac{1}{\sqrt{6}}(\ket{\downarrow_1 \uparrow_2 \downarrow_3} + \ket{ \uparrow_1 \downarrow_2 \downarrow_3} - 2\ket{\downarrow_1 \downarrow_2 \uparrow_3 }) ,
   \end{split}
\end{equation}
as well as
 \begin{equation}\label{eq:Quartets_Q}
  \begin{split} 
   & \ket{1,\tfrac{3}{2},\tfrac{3}{2}} = \ket{\uparrow_1 \uparrow_2 \uparrow_3 },\\
  & \ket{1,\tfrac{3}{2},\tfrac{1}{2}} = \frac{1}{\sqrt{3}}(\ket{\uparrow_1 \uparrow_2 \downarrow_3 }+\ket{\downarrow_1 \uparrow_2  \uparrow_3 } +\ket{ \uparrow_1  \downarrow_2 \uparrow_3 }), \\ 
  & \ket{1,\tfrac{3}{2},-\tfrac{1}{2}} = \frac{1}{\sqrt{3}}(\ket{\downarrow_1 \downarrow_2 \uparrow_3  }+\ket{\uparrow_1 \downarrow_2  \downarrow_3 } +\ket{ \downarrow_1  \uparrow_2 \downarrow_3 }), \\
  & \ket{1,\tfrac{3}{2},-\tfrac{3}{2}} = \ket{\downarrow_1 \downarrow_2 \downarrow_3 }.
   \end{split}
\end{equation}
The same construction applies to a right trimer $\Delta'$.
We note that 
for an isolated isosceles trimer with $J_1=J_2$, the eigenstates of the Heisenberg Hamiltonian  
consists of the two twofold degenerate doublets 
$D_0$ (Eq.~\eqref{eq:Dubletts_D0}) with energy $E_{D_0}=-\tfrac{3}{4}J_3$,
and 
$D_1$ (Eq.~\eqref{eq:Dubletts_D1}) with energy $E_{D_1}=\tfrac{1}{4}J_3-J_1$,
and the fourfold degenerate quartet 
$Q$ (Eq.~\eqref{eq:Quartets_Q}) with energy $E_{Q}=\tfrac{1}{2}J_1+\tfrac{1}{4}J_3$
. It therefore depends on  the coupling ratio, which doublet forms the ground state. For $J_1=J_2<J_3$ ($J_1=J_2>J_3$),  $D_0$ ($D_1$) has the lowest energy. For $J_1=J_2=J_3$, the two doubles $D_0$ and $D_1$ become degenerate and the ground state is fourfold degenerate.
In all above cases, the quartet has the highest energy. We refer to Refs.~\cite{Weber2022B} for details on the QMC SSE algorithm in the trimer basis.

\section{Sign problem analysis}\label{Sec:Sign}

For the spin-1/2 Heisenberg model on the star lattice, QMC simulations can be performed without any sign problem only in a few limiting cases: (i) when $J_d=0$ there is no sign problem in the trimer basis,  (ii) if at least one of the three trimer couplings vanishes, there is no sign problem in the site basis, since the system is bipartite under this condition, (iii) in the limit of vanishing trimer coupling there is no sign problem in both the site and the dimer basis. For the generic case, however, there is a QMC sign problem for each of the three considered computational basis,  and we therefore examine in more detail the average sign, $\langle \mathrm{sign}\rangle$, for the different algorithms in order to compare their performance~\cite{Weber2022A}. For this purpose, it is convenient to introduce an angular parameter $\phi$, in terms of which
\begin{equation}
J_d=\bar{J}\cos(\phi),\quad J_i=\bar{J}\sin(\phi),
\end{equation}
for $i=1,2,3$, i.e., we consider the case of equal intratrimer couplings, and $\bar{J}$ quantifies the overall interaction scale. This parameterization allows us to conveniently tune from the limit of isolated $J_d$-dimers for $\phi=0$ to isolated trimers for $\phi=\pi/2$.
\begin{figure}[t]
    \centering
    \includegraphics[width=0.5\textwidth]{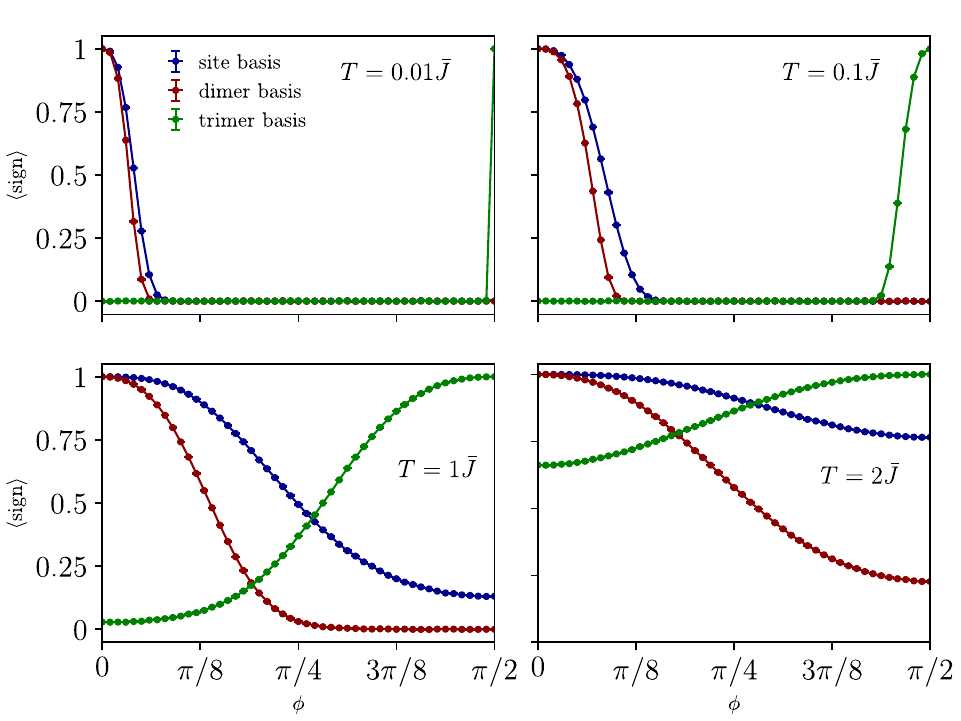}
    \caption{Average QMC sign, $\langle\mathrm{sign}\rangle$, of the spin-1/2 Heisenberg model on the star lattice ($L=4$) as a function of $\phi$ for various temperatures $T$, specified in units of $\bar J$.}
    \label{Fig:Sign}
\end{figure}
The values of $\langle \mathrm{sign}\rangle$ for the different algorithms is shown in Fig.~\ref{Fig:Sign} as a function of $\phi$ at four selected temperatures $T$, specified in units of $\bar{J}$. For a completely sign-free QMC simulation, the value of $\langle \mathrm{sign}\rangle=1$, independently of temperature. This behavior is observed in Fig.~\ref{Fig:Sign}  in both of the aforementioned limits. Beyond these limits, all methods exhibit a sign-problem, as indicated by a value of $\langle \mathrm{sign}\rangle<1$. Furthermore, upon reducing $T$, the averaged sign gets strongly reduced, where a value below about $0.01$ renders efficient QMC sampling unfeasible due to a significant increase in computational resources needed in order to reduce the statistical uncertainly. More relevant than the actual value of $\langle \mathrm{sign}\rangle$ within this inaccessible regime is in fact its behavior as a function of $\phi$ for the different algorithms. In order to discuss this in more detail, we focus on the case $T=\bar{J}$ in Fig.~\ref{Fig:Sign}. For dominant intratrimer couplings, i.e., for $\phi>\pi/4$, the trimer basis algorithm is seen to be favorable in terms of the value of $\langle \mathrm{sign}\rangle$. This is certainly expected from the limiting behavior as $\phi$ approaches $\pi/2$, where the system decouples into uncoupled trimers. 
Correspondingly, we find that upon approaching the other limit, $\phi\rightarrow 0$, the average sign drops strongly for this algorithm, while the average sign for both the site and the dimer basis increase. 
Interestingly, we find that for all values of $\phi$, the site-based algorithm exhibits a larger value of $\langle \mathrm{sign}\rangle$ than the dimer-based algorithm. Indeed, even in the regime of dominant $J_d$ coupling, where the formation of singlet correlations on the $J_d$ dimer bonds strongly increases, finite residual interdimer correlations remain, which apparently are described less efficiently in the dimer basis. It furthermore turns out that for the homogeneous case with $\phi=\pi/4$, i.e., $J_d=J_1=J_2=J_3$, the site and the trimer basis perform similarly in terms of the average sign, with only a slight advantage for the site-based algorithm. In view of the above, this is not too surprising, given that in the homogeneous case the ground state was found to be dominated by the formation of singlets on the  dimer bonds~\cite{Richter2004,richter2004starlattice}. Based on the above analysis, we  therefore used the trimer-based  algorithm for the regime of dominant trimer coupling, and otherwise used the site-based QMC algorithm for obtaining  the thermodynamic results presented below. 

\section{Results}\label{Sec:Results}

We now discuss the thermodynamic properties of the spin-1/2 Heisenberg model on the star lattice within the various regimes of the interaction strengths. 

\subsection{QMC results for dominant dimer coupling}\label{Sec:WeakIntra}

We first consider the case that the dimer  coupling $J_d$ dominates over the three trimer couplings $J_1$, $J_2$ and $J_3$. In this regime, the ground state is characterized by the formation of spin singlets on the dimer bonds. Indeed, this physics was previously established for the balanced case, where all couplings are of equal strengths, $J_d=J_1=J_2=J_3$~\cite{Richter2004,richter2004starlattice}, 
as quoted in the introduction. In the following, we will denote by $J$ the value of the couplings in the balanced case, and by $J_t$ the value of the $J_i$, $i=1,2,3$ in cases where they are all equal --  in the balanced case, considered in Refs.~\cite{Richter2004,richter2004starlattice}, thus $J_d=J_t=J$ holds. 

\begin{figure}[t]
    \centering

    \vspace{-0.7cm}
    
    \includegraphics[width=0.5\textwidth]{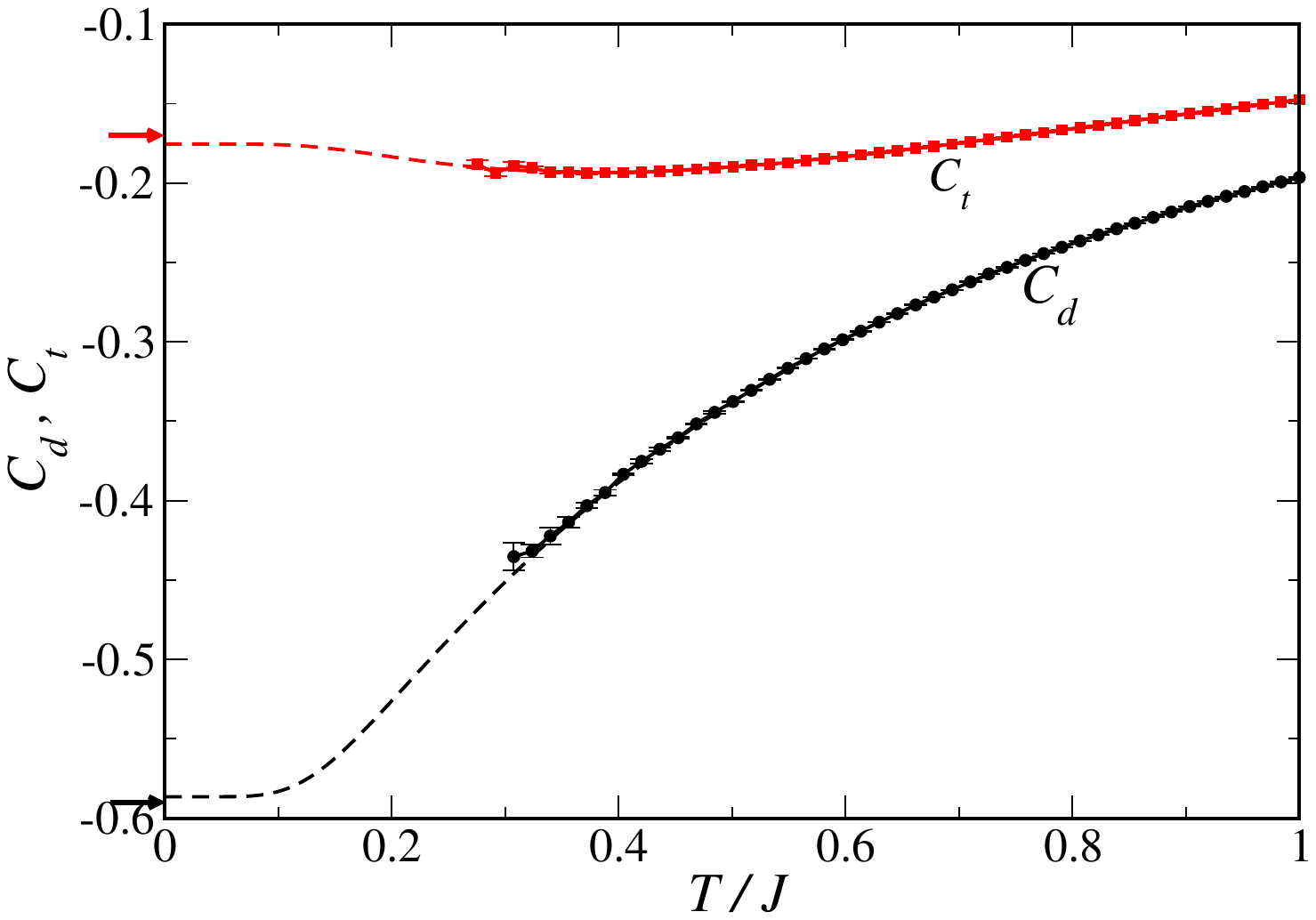}
    \caption{Temperature dependence of the  nearest-neighbor correlations $C_d=\langle \mathbf{S}_i \cdot \mathbf{S}_j\rangle_{d}$ and  $C_t=\langle \mathbf{S}_i \cdot \mathbf{S}_j\rangle_{t}$ along a dimer and trimer bond, respectively, for the homogeneous spin-1/2 Heisenberg model on the star lattice as obtained from QMC ($N=24$, symbols) and ED ($N=18$, dashed lines). $T=0$ ED values from Ref.~\cite{Richter2004} ($N=36$) are indicated by arrows. } 
    \label{Fig:bondstrength}
\end{figure}

To illustrate the dominance of the dimer couplings  in the balanced case, we show in Fig.~\ref{Fig:bondstrength} the two non-equivalent nearest-neighbor spin-spin correlations $C_d=\langle \mathbf{S}_i \cdot \mathbf{S}_j\rangle_{d}$  and $C_t=\langle \mathbf{S}_i \cdot \mathbf{S}_j\rangle_{t}$ along a dimer and trimer bond, respectively. Here,  finite-$T$ results from QMC simulations for  $N=24$ ($L=2$) 
and ED for $N=18$ are compared to the ground state values  
for $N=36$ reported in Ref.~\cite{Richter2004} (due to the sign problem, we cannot reach towards lower temperatures in these QMC simulations). The correlations along the dimer bonds are clearly seen to dominate over those on the trimer bonds. Furthermore, no significant finite-size effects are observed in these quantities, as expected  from the  rather strong dimerization of the system on the dimer bonds. We  note a slight non-monotonous behavior in the strength of the  nearest-neighbor spin-spin correlations along the trimer bonds  upon approaching  towards the ground state: At finite temperatures, the weakening of the correlations on the dimer bonds with respect to the ground state thus leads to a slight initial  enhancement of the correlations on the trimer bonds before eventually thermal fluctuations weaken all correlations altogether. 

\begin{figure}[t!]
    \centering

    \vspace{-0.7cm}
    
    \includegraphics[width=0.45\textwidth]{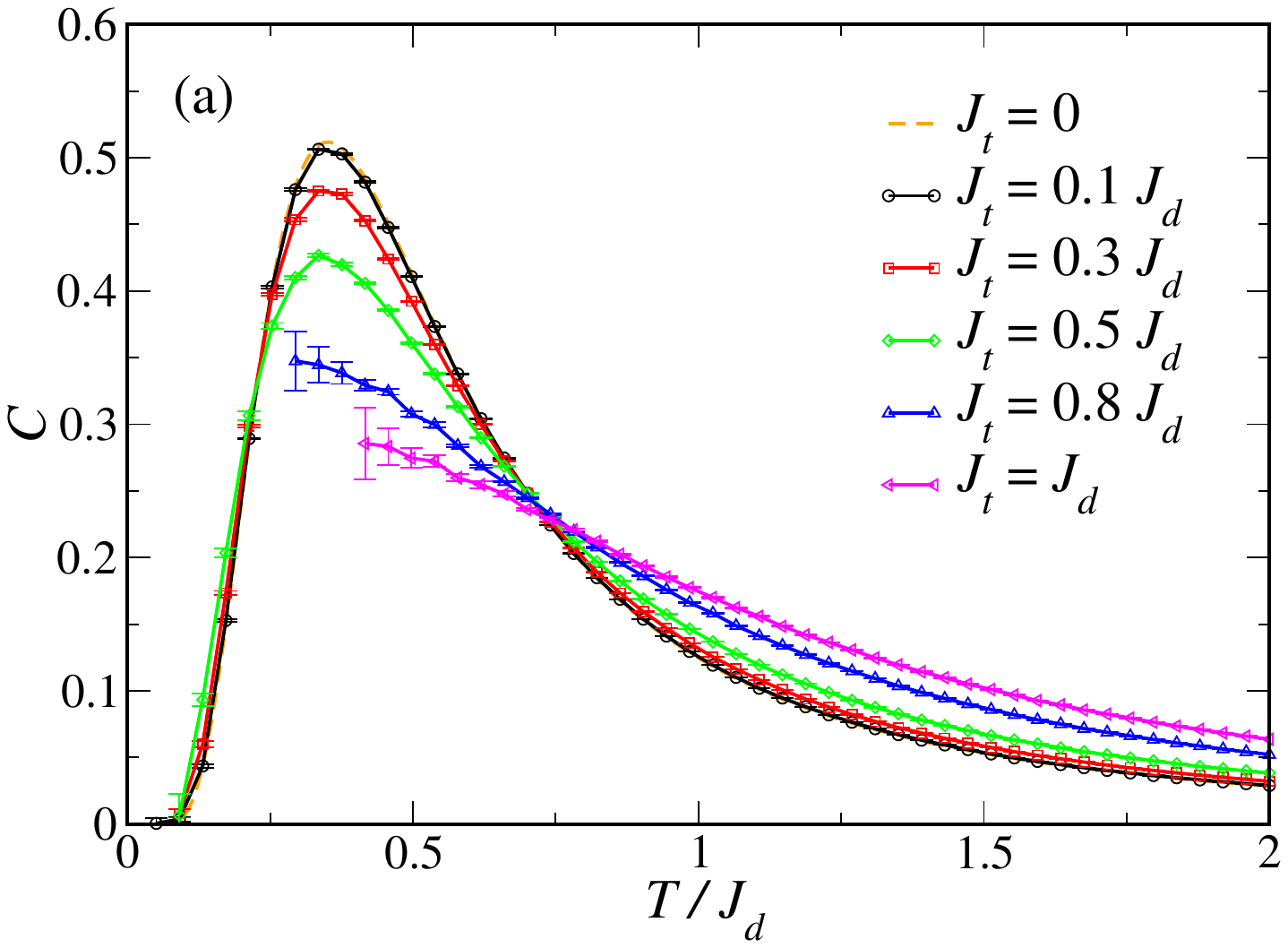}

    \vspace{-1cm}
    
    \includegraphics[width=0.45\textwidth]{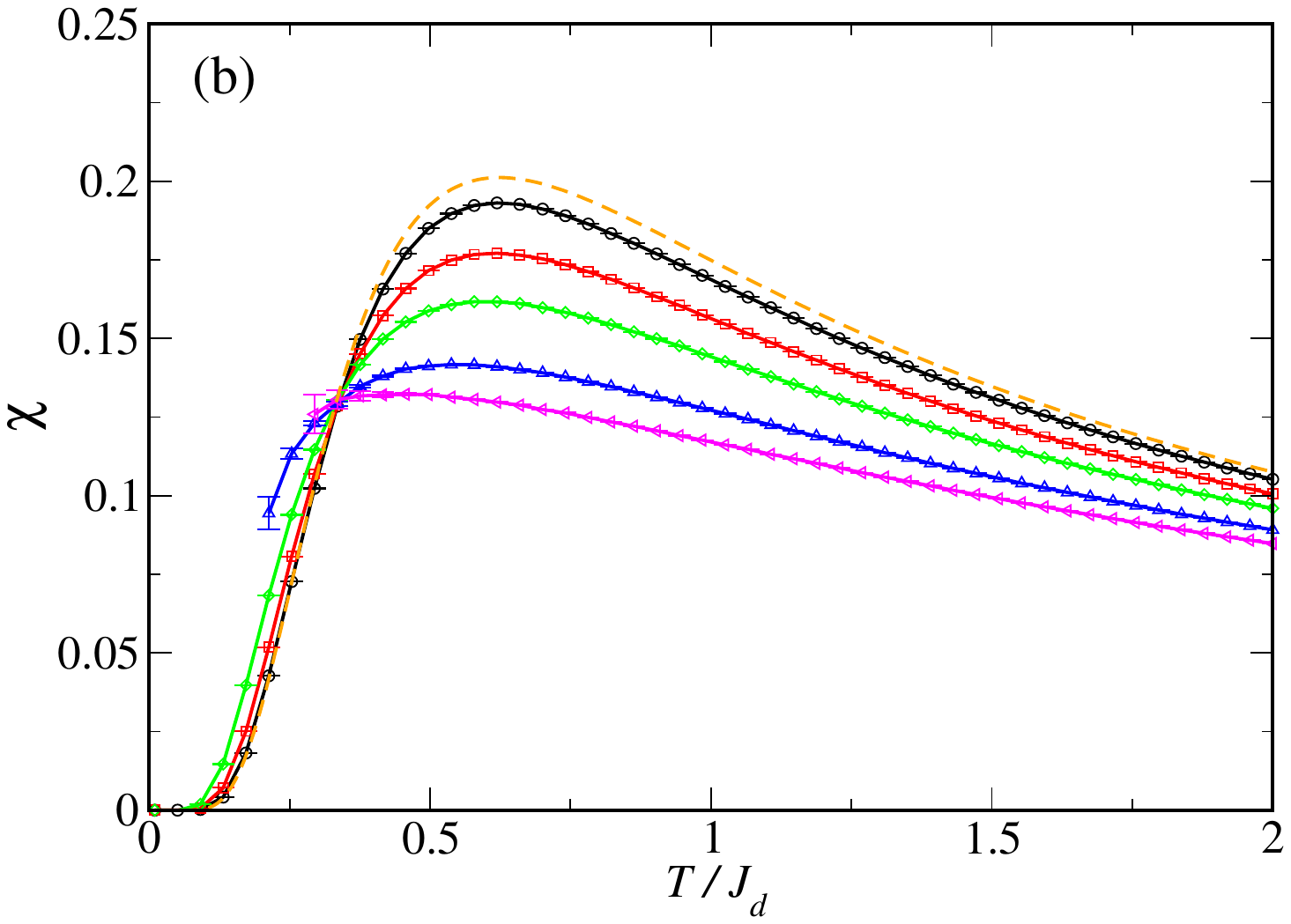}

    \vspace{-1cm}
    
    \includegraphics[width=0.45\textwidth]{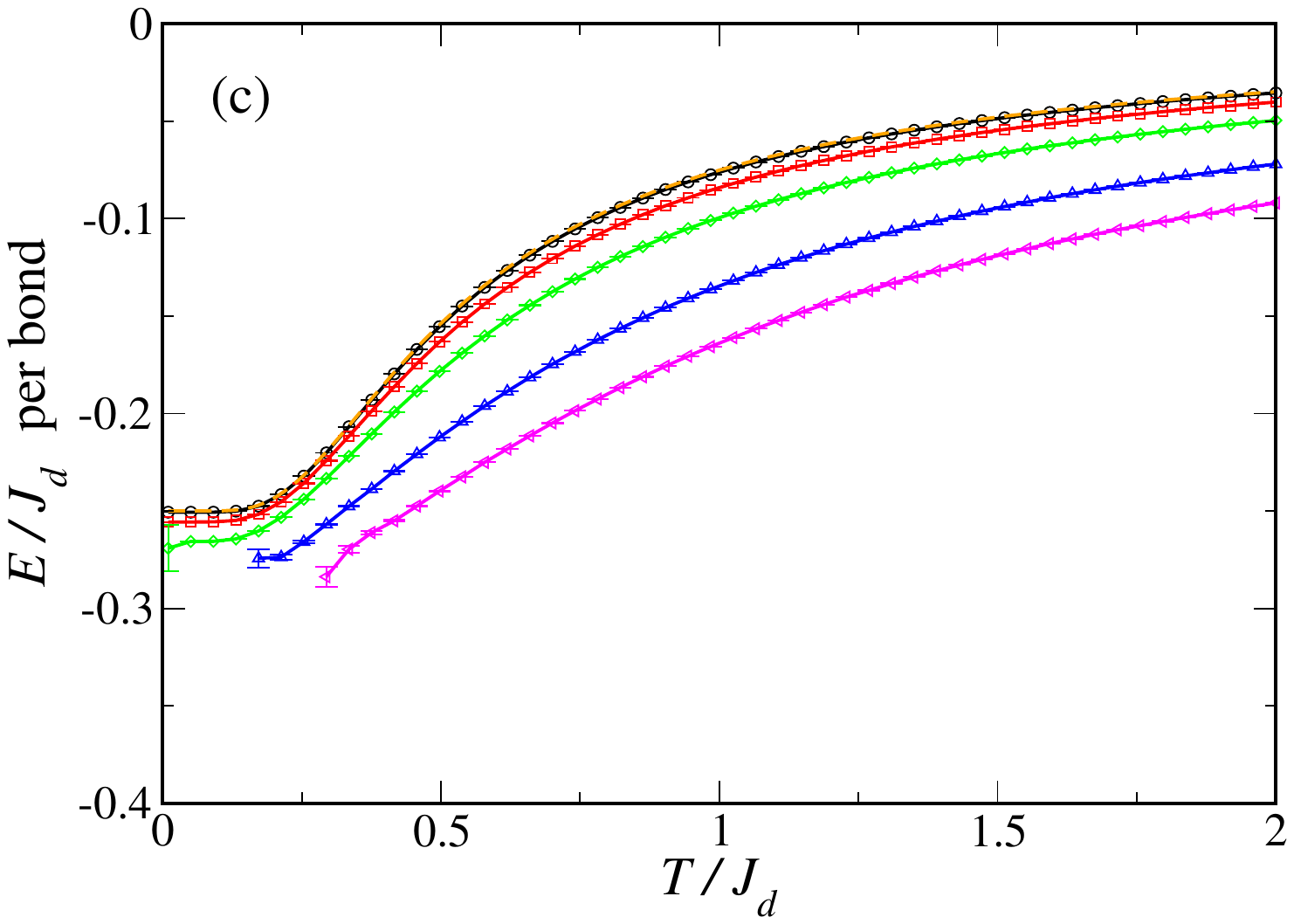}   
    \caption{Temperature dependence of (a) the specific heat $C$, (b) the uniform susceptibility $\chi$, and (c) the energy $E$ per bond of the spin-1/2 Heisenberg model on the star lattice as obtained from QMC simulations ($L=2$) for different values of the trimer coupling $J_t$ (data for  isolated dimers at $J_t=0$ are from ED for $N=2$).}
    \label{Fig:stronginter}
\end{figure}

In Ref.~\cite{Richter2004}, a large spin gap (singlet-triplet gap) of $\Delta/J=0.3809$ was identified for the balanced case. Such sizeable spin gaps relate to a correspondingly strong suppression in thermodynamic response functions at low temperatures in the regime of dominant dimer couplings.
This is exhibited by our QMC data for the temperature dependence of both the specific heat $C$ and the uniform magnetic susceptibility $\chi$  shown in Fig.~\ref{Fig:stronginter}. Due to the sign problem, we can only access the relevant low-$T$ regime for values of $J_t/J_d$ below about $0.5$. The bottom panel of Fig.~\ref{Fig:stronginter} also shows the temperature dependence of the energy $E$, which approaches closer to the single-dimer $(J_t=0)$ 
behavior upon decreasing $J_t$, as expected. For the homogeneous case, 
a ground state energy per bond of $E/J=-0.3093$ 
was reported in Ref.~\cite{Richter2004}, and our data for varying $J_t/J_d$ is in accord with an approach towards this value, even though we certainly cannot access the low-$T$ regime for $J_t/J_d$ beyond about $0.5$, again due to the QMC sign problem. However, overall we obtain a consistent picture that in the regime of dominant dimer coupling the low-$T$ thermodynamic properties are characterized by the dominant singlet formation on the dimer bonds and an associated sizeable spin gap. Finally, we note that the data in both Fig.~\ref{Fig:stronginter} (a) and (b) exhibit apparent crossing points in $C(T)$ and $\chi(T)$ for different ratios $J_t/J_d$ at two (different) temperatures, $T\approx 0.7J_d$ and $T\approx 0.4J$, respectively. Such isosbestic  points have  been observed and discussed for response functions in various systems~\cite{Vollhardt97} and occur, e.g., also for the Hubbard model in different dimensions~\cite{chandra99}. Upon closer inspection, the crossing points seen in Fig.~\ref{Fig:stronginter} however actually exhibit a weak systematic drift instead of  actual isosbestic behavior. 
\\

Unfortunately, our QMC approach does not allow us to explore the parameter regime where the 
trimer couplings $J_i$ are of order but larger than $J_d$, since in this regime, the sign problem is 
rather severe in all the computational bases that we considered. 
Therefore, we use the  FTLM to calculate $C$ and $\chi$ down to low temperatures for this
parameter regime. We discuss these data in the next section. 
In the regime of strong trimer couplings, i.e., for weak dimer coupling, $J_d \ll J_1,J_2,J_3$, one can 
explore the magnetic 
properties of the spin-1/2 Heisenberg model on the star lattice based on perturbation theory, in terms of effective 
Hamiltonians for the trimer total spin degrees of freedom. 
We turn to discuss this approach in Sec.~\ref{Sec:EffHam}. 

\subsection{FTLM results for (almost) balanced dimer and trimer couplings}
\label{Sec:Balanced}
As discussed in Sec.~\ref{Sec:Sign} and \ref{Sec:WeakIntra},
the QMC sign problem becomes increasingly severe upon approaching the balanced case $J_t=J_d$.
However, from a theoretical point of view it is of particular interest to
study the case $J_t=J_d=J$, i.e., the pure Archimedean-lattice model. Namely, besides the celebrated kagome lattice the star lattice is the only other Archimedean lattice for which the 
interplay of quantum fluctuations and
frustration prevents ground-state
magnetic ordering  for the spin-$1/2$ Heisenberg antiferromagnet.  Another example of a non-magnetic quantum
ground state, which has attracted much attention
recently, is provided by the spin-$1/2$ Heisenberg antiferromagnet on the square-kagome
lattice (which is not an Archimedean lattice)~\cite{richter2009squago,Sakai2013,Rousochatzakis2015,Lugan2019,squago_NatCom2020,Iqbal2021,Vasiliev2022,schmoll2022tensor}.
Several studies already examined the thermodynamic properties of the  spin-$1/2$ Heisenberg antiferromagnet
on the kagome
\cite{Lhuillier_thermo_PRL2000,Bernu2005,Rigol2007,Shimokawa2016,Xi-Chen2018,kago42} and, more recently, the 
square-kagome lattice \cite{squago_TD_2022,squago_j1j2_2023}. 
In particular,  Refs.~\cite{kago42} and \cite{squago_TD_2022} also used the FTLM. It is thus interesting to compare our star-lattice FTLM data with those for the kagome and square-kagome lattices.  

\begin{figure}[t]
    \centering

    \vspace{-0.7cm}
    
    \includegraphics[width=0.47\textwidth]{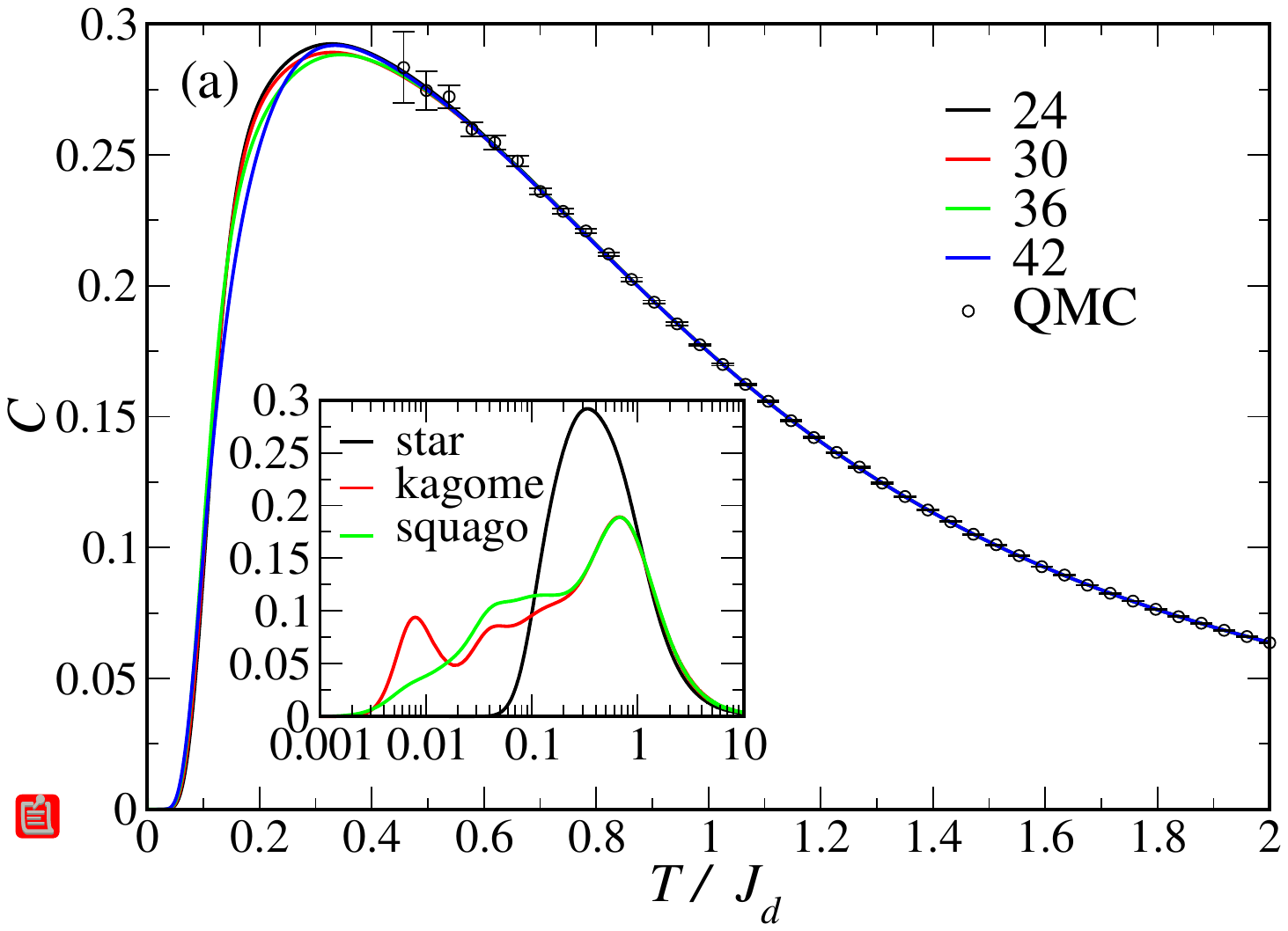}

    \vspace{-1cm}
    
    \includegraphics[width=0.47\textwidth]{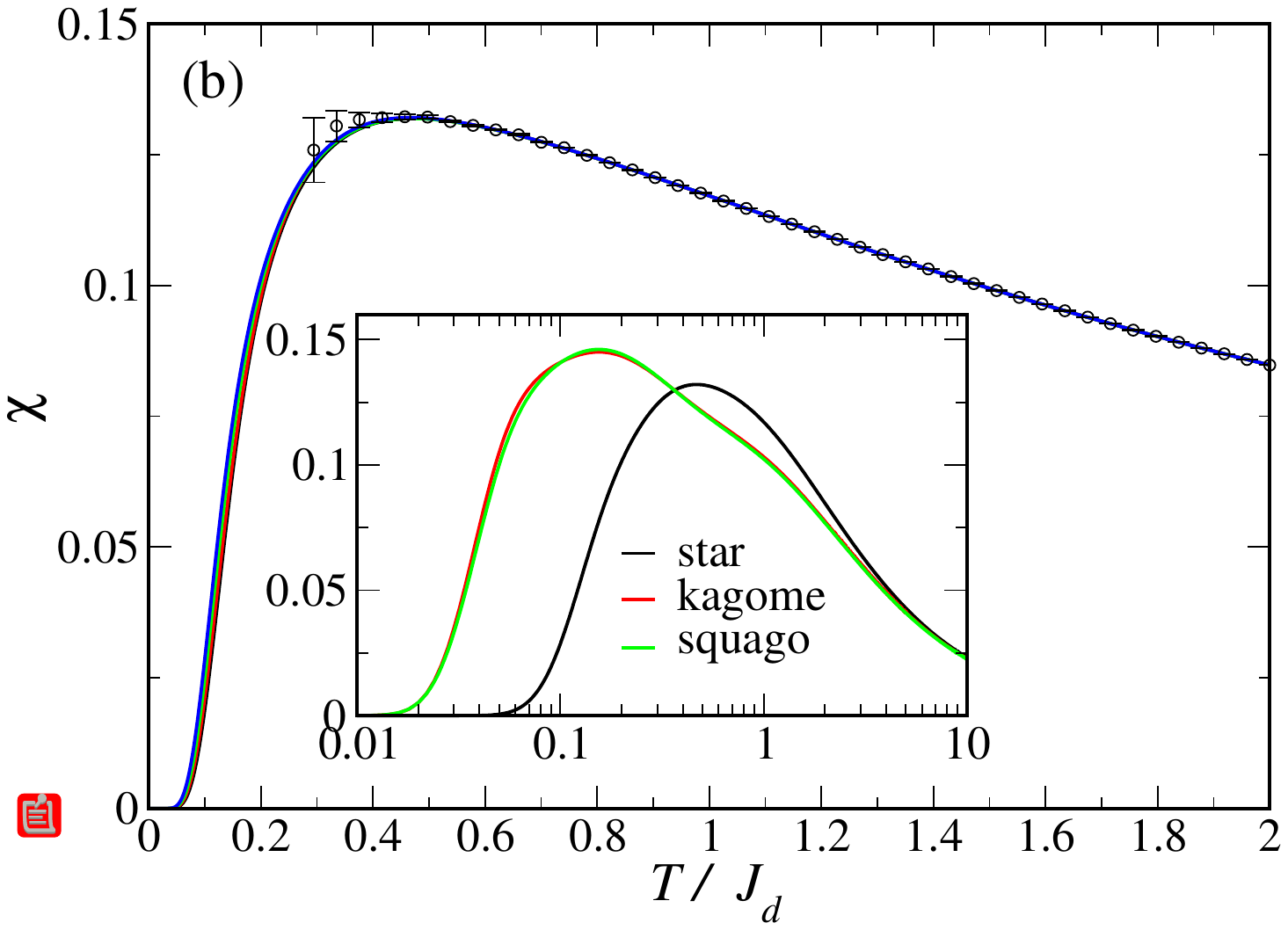}
    \caption{Main panels: Temperature dependence of (a) the specific heat $C$ and (b) the 
    uniform susceptibility $\chi$ of the spin-1/2 Heisenberg model on the balanced star lattice
    ($J_d=J_t=1$) as 
    obtained by FTLM for finite lattices of N=24,30,36,42 sites and from QMC
    simulations.
    Inset: Comparison of the temperature dependence (logarithmic scale) of (a) the specific heat $C$ and (b) the uniform susceptibility
    $\chi$ of the spin-1/2 Heisenberg model on the star, kagome
    \cite{kago42} and 
    square-kagome (squago) \cite{squago_TD_2022} lattices ($N=42$).}
    \label{Fig:1_1_FTLM}
\end{figure}

We present the temperature dependence of the specific heat $C$ and the uniform susceptibility $\chi$ in Fig.~\ref{Fig:1_1_FTLM}. 
Obviously, there are only very weak finite-size effects, which can be attributed to the strong 
dimerization that leads to a very short magnetic correlation length. 
Only around the maximum in the specific heat do the $C(T)$ curves for different system sizes deviate slightly from each other.
In the temperature region where reliable QMC data are available, they agree very well
with the FTLM data.

A comparison with the corresponding data for the kagome and square-kagome lattice models
(shown in the  insets of Fig.~\ref{Fig:1_1_FTLM})
demonstrates that the 
thermodynamics of the star-lattice 
Heisenberg antiferromagnet is significantly
different from that of the kagome and square-kagome lattices.
In particular, 
the pronounced dimerization of the star lattice model leads to a large spin gap (singlet-triplet gap),
and hence the maximum in $\chi(T)$ is located at much higher temperature than in the other two models.  Furthermore, 
by contrast to the kagome and square-kagome models there are no low-lying
singlet excitations within
the spin gap. Thus, 
the extra low-temperature features (maximum/shoulder) in $C(T)$ are only
present for the kagome and square-kagome lattices. 

\begin{figure}[t]
    \centering

    \vspace{-0.7cm}
    
    \includegraphics[width=0.47\textwidth]{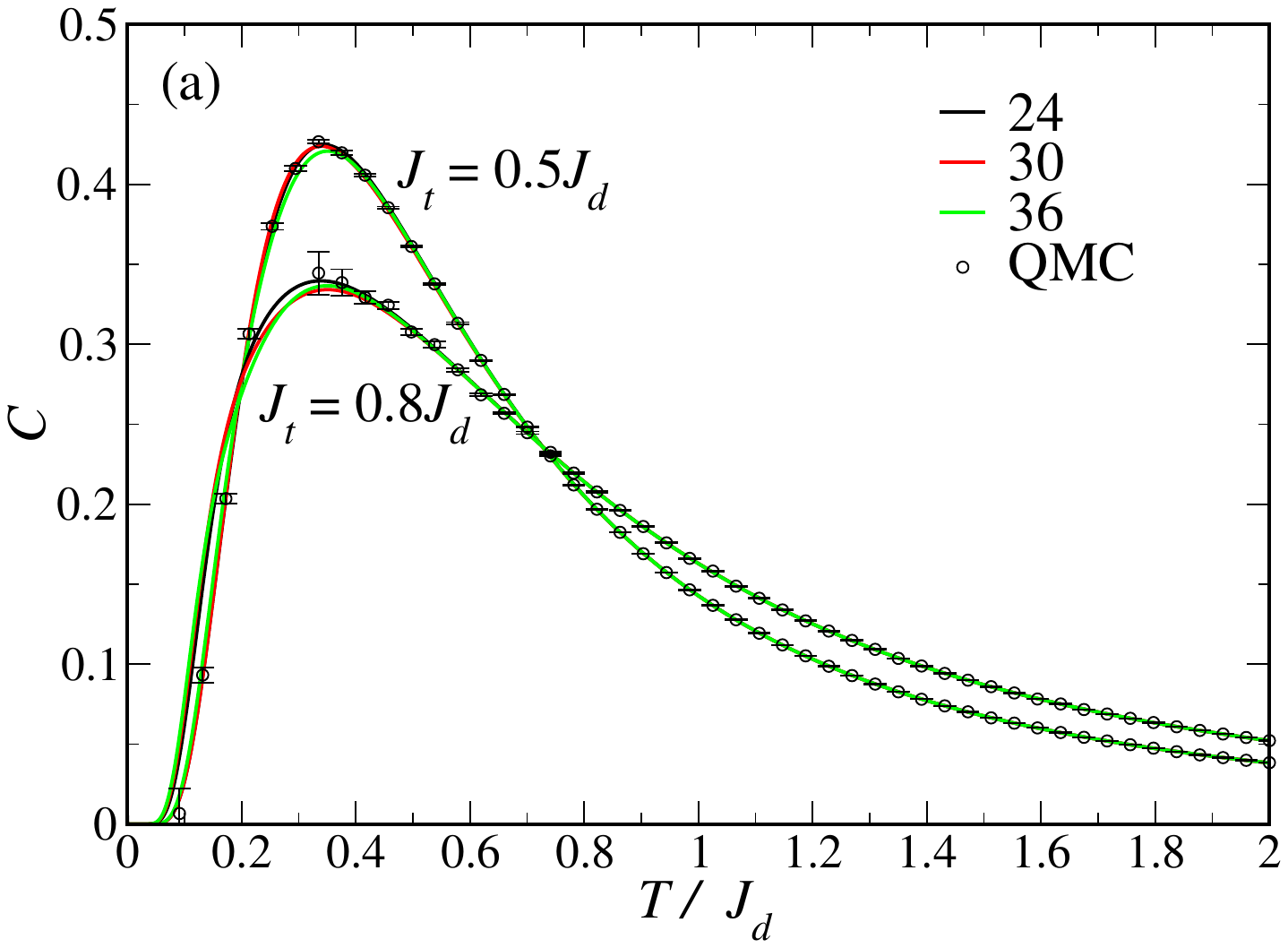}

    \vspace{-1cm}
    
    \includegraphics[width=0.47\textwidth]{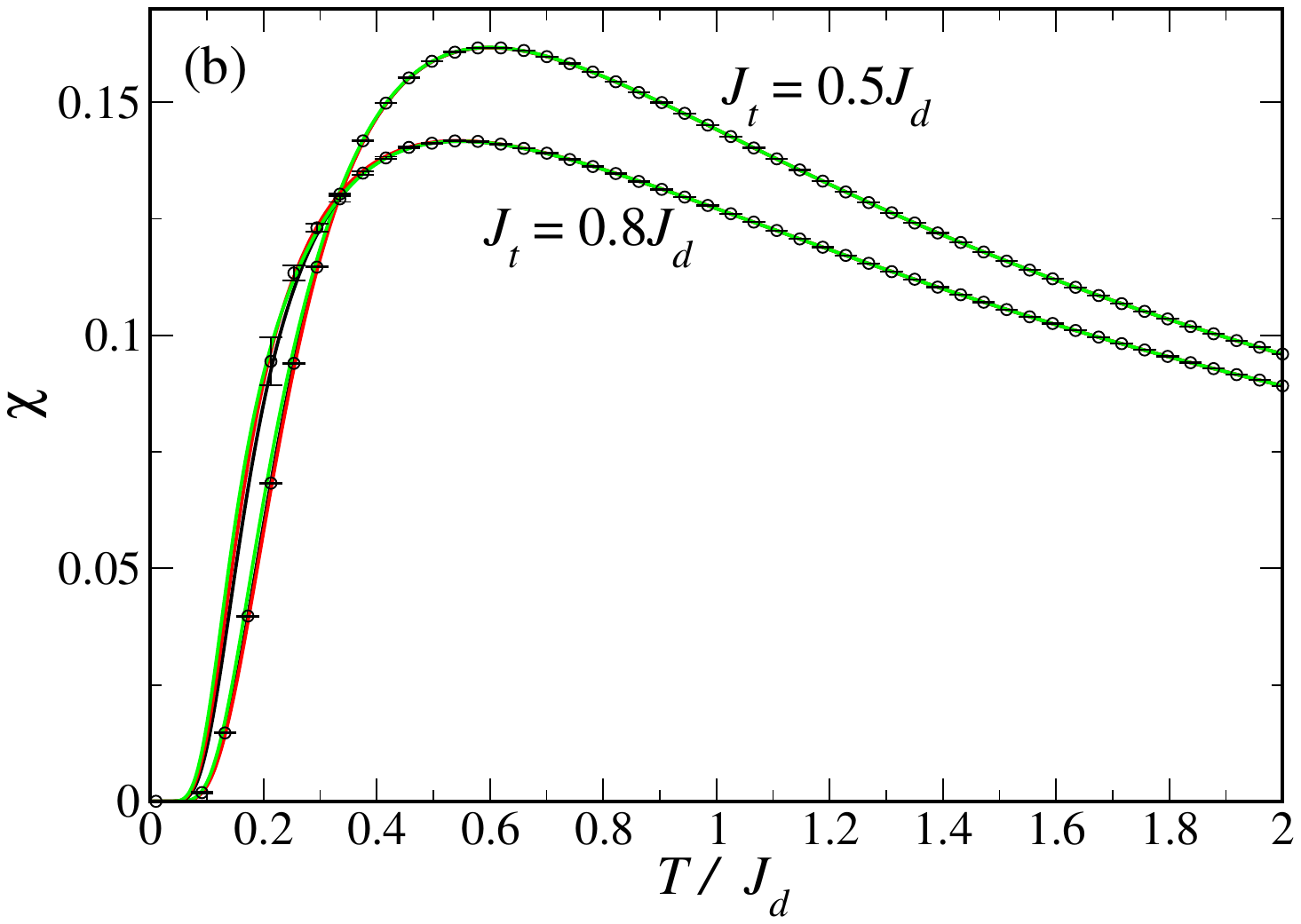}
    \caption{Temperature dependence of (a) the specific heat $C$ and (b) the 
    uniform susceptibility $\chi$ of the spin-1/2 Heisenberg model on the star lattice
    with $J_d=0.5J_t$ and $J_d=0.8J_t$ as 
    obtained by FTLM for finite lattices of $N=24,30,36$ sites and from QMC
    simulations.
}
    \label{Fig:1_0.5_0.8_FTLM}
\end{figure}

Finally,  we compare FTLM and QMC data for $J_t=0.5J_d$  and $J_t=0.8J_d$
in Fig.~\ref{Fig:1_0.5_0.8_FTLM}.
The general shape of the temperature profiles is very similar to that for
$J_t=J_d$, where the position and height of the maxima in  $C(T)$ and $\chi(T)$ naturally
depend on the ratio $J_t/J_d$.
As can be seen for $J_t=0.5J_d$ the agreement between the QMC  and the
FTLM data is excellent down to very low temperature, reconfirming the accuracy of the FTLM approach also within this temperature range.

\subsection{Effective Hamiltonians for weak dimer coupling}\label{Sec:EffHam}

In order to describe the low-temperature physics of the spin-1/2 Heisenberg model on the star lattice for  weak dimer coupling, an effective low-energy Hamiltonian can be derived from first-order Brillouin-Wigner perturbation theory, analogous to the triangle square lattice case~\cite{Wang2001, Weber2022A}.
In this section, we first present the resulting effective models. 

\subsubsection{Isosceles trimer couplings}

Away from the special point  $J_1 = J_2 = J_3$, which will be detailed below,  the low-energy states of each trimer are formed by a doublet of states, forming an effective spin-1/2 degree of freedom on each trimer, given by $\mathbf{s}_{\Delta} = P\mathbf{S}_{\Delta} P $ (where $P$ is the projector to the lowest energy doublet) and $\mathbf{s}_{\Delta'} = P\mathbf{S}_{\Delta'} P $.
These effective trimer spins are arranged on a honeycomb lattice (cf. Fig.~\ref{Fig:lattice}), which is bipartite. A (weak) finite dimer interaction $J_d$ leads to an antiferromagnetic coupling between these trimers by effective Heisenberg interactions, with strengths that depend on the orientation of the dimer bonds, as detailed below. 
Due to the bipartite nature of the honeycomb lattice, this can lead to  antiferromagnetic order among the effective trimer spins in the ground state. However, it is also possible that a quantum disordered state forms in case  a particular effective coupling in $H_\mathrm{eff}$ dominates over the others. As shown below, both situations are indeed realized. 
The form of the effective Hamiltonian is given by 
\begin{equation}
H_\textrm{eff} = \sum_{\Vec{r}}  J_\textrm{eff}^{(1)}\mathbf{s}_{\Delta'}^{\Vec{r}} \cdot \mathbf{s}_{\Delta} ^{\Vec{r} + \Vec{a_1}}  +J_\textrm{eff}^ {(2)} \mathbf{s}_{\Delta'}^{\Vec{r}} \cdot \mathbf{s}_{\Delta} ^{\Vec{r} + \Vec{a_2}}
+ J_\textrm{eff}^{(3)}\mathbf{s}_{\Delta} ^{\Vec{r}}\cdot \mathbf{s}_{\Delta'}^{\Vec{r}} ,
\end{equation}
up to a irrelevant constant term. 
Analytical expressions of the effective couplings $J_\textrm{eff}^{(1)}$, $J_\textrm{eff}^{(2)}$ and $J_\textrm{eff}^{(3)}$ will first be discussed below for the specific case $J_1 = J_2 \neq J_3$, and later the general case will be considered. 

\subsubsection*{The case $J_3 > J_1 = J_2$}

\begin{figure}[t]
    \centering
    \includegraphics[width=0.5\textwidth]{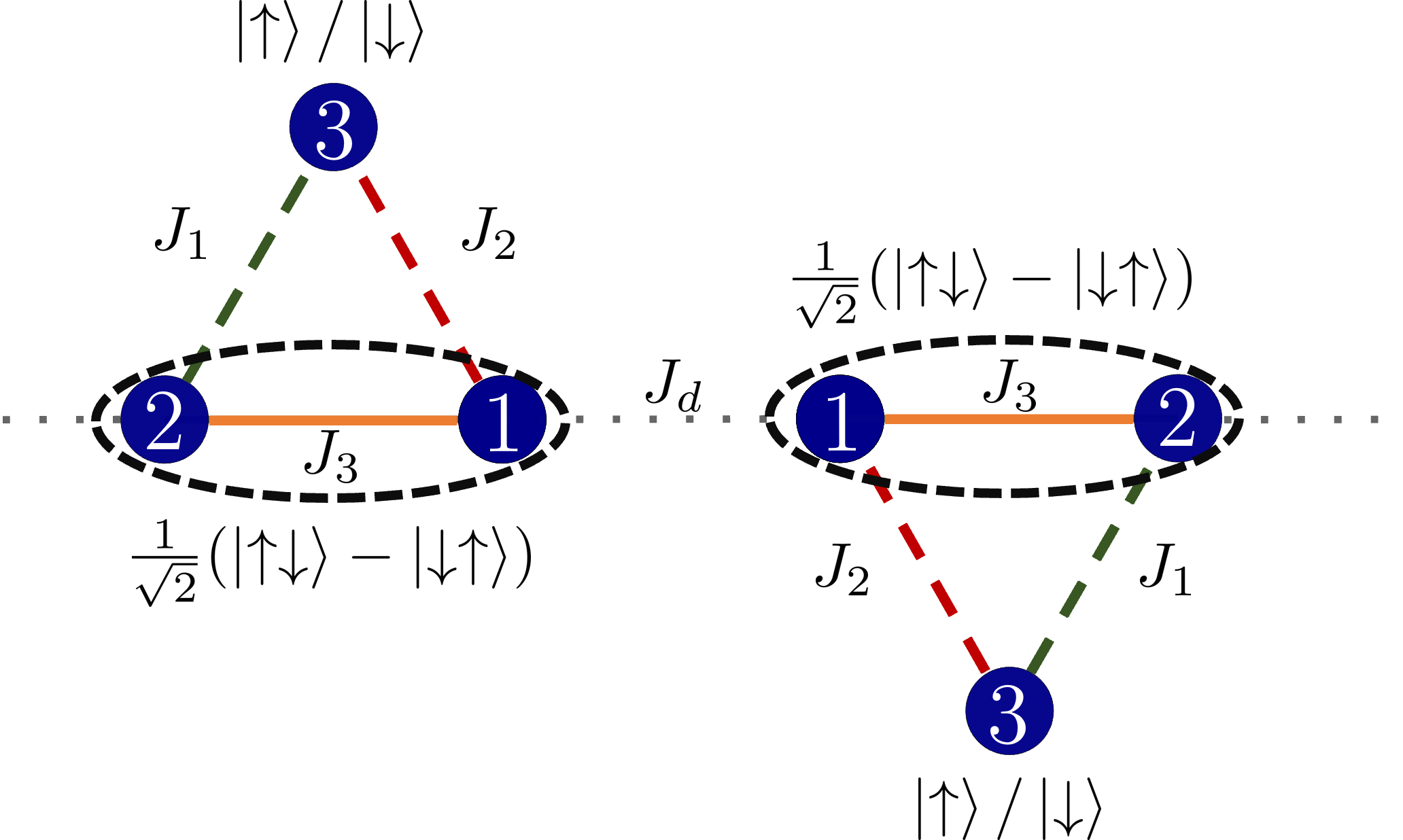}%
    \caption{Illustration for the determination of the effective Hamiltonian along an intertrimer bond in the $\Vec{a}_1$  direction  for $J_3 > J_1 = J_2$. In each trimer, the ground state consists of a direct product of the singlets among the two spins interacting via $J_3$ and an $\ket{\uparrow}_3$ or $\ket{\downarrow}_3$ state of the third spin.}
    \label{fig:BondIllustration}
\end{figure}

In this case the ground state of each isolated trimer consists of the doublet $D_0$, i.e., the two states
\begin{eqnarray}
\ket{+} &=&\ket{0,\tfrac{1}{2},\tfrac{1}{2}} = \frac{1}{\sqrt{2}}(\ket{\uparrow_1 \downarrow_2} - \ket{ \downarrow_1 \uparrow_2})\otimes \ket{\uparrow_3},  \nonumber \\
\ket{-} &=&\ket{0,\tfrac{1}{2},-\tfrac{1}{2}} = \frac{1}{\sqrt{2}}(\ket{\uparrow_1 \downarrow_2} - \ket{ \downarrow_1 \uparrow_2})\otimes \ket{\downarrow_3}.\nonumber 
\end{eqnarray}
To derive the effective Hamiltonian, we consider two neighboring trimers, and for each dimer bond evaluate the matrix elements of the spin-spin coupling term corresponding to that bond in the above basis. From the resulting $4\times 4$ matrix, the effective Hamiltonian corresponding to the considered intertrimer bond is then readily determined. 
We obtain $J_\textrm{eff}^{(3)}=J_d$, while the other effective couplings vanish, $J_\textrm{eff}^{(1)}=J_\textrm{eff}^{(2)}=0$, to leading order in perturbation theory. The effective model in this case thus describes decoupled dimers of the effective trimer spins. Since this effective dimer coupling is positive, the ground state consists of a product state of effective dimer singlets.  
The fact that the effective coupling vanishes for the two bonds along the directions of  $\Vec{a}_1$ and $\Vec{a}_2$ can be understood by examining the ground states of the uncoupled trimer. These consist of a product of the singlet state $\ket{S} = \frac{1}{\sqrt{2}}(\ket{\uparrow_1 \downarrow_2} - \ket{ \downarrow_1 \uparrow_2})$ resulting from the coupling between the spins $\mathbf{S}_{1,\Delta}$ and $\mathbf{S}_{2,\Delta}$, with either the $\ket{\uparrow_3}$ or the $\ket{\downarrow_3}$ state for the third spin $\mathbf{S}_{3,\Delta}$. The two states $\ket{+}$ and $\ket{-}$ therefore differ only by a flip of the third spin. 
The off-diagonal matrix elements of the effective Hamiltonian therefore vanish because $\braket{\uparrow_3|\downarrow_3} = 0$.
The diagonal elements also vanish along the bonds in the $\Vec{a}_1$ and $\Vec{a}_2$ directions, since from the antisymmetry of the singlet states it follows that  
$ \bra{S} \mathbf{S}_{k,\Delta} \cdot \mathbf{S}_{k,\Delta'}\ket{S}=0$
for both $k = 1,2$ (cf. Fig.~\ref{fig:BondIllustration} for an explicit illustration for the case $k=1$).

\subsubsection*{The case  $J_3 < J_1 = J_2$}

In this case the ground state of each isolated trimer consists of the doublet $D_1$, i.e., the two states
\begin{eqnarray}
 \ket{1,\tfrac{1}{2},\tfrac{1}{2}} &=& \frac{1}{\sqrt{6}}(\ket{\uparrow_1 \downarrow_2 \uparrow_3} + \ket{ \downarrow_1 \uparrow_2 \uparrow_3} - 2\ket{\uparrow_1 \uparrow_2 \downarrow_3 }),\nonumber \\  
 \ket{1,\tfrac{1}{2},-\tfrac{1}{2}} &=& \frac{1}{\sqrt{6}}(\ket{\downarrow_1 \uparrow_2 \downarrow_3} + \ket{ \uparrow_1 \downarrow_2 \downarrow_3} - 2\ket{\downarrow_1 \downarrow_2 \uparrow_3 }),\nonumber
\end{eqnarray}
and from first-order perturbation theory we now obtain $J_\textrm{eff}^{(1)}=J_\textrm{eff}^{(2)}=\tfrac{4}{9}J_d$ and $J_\textrm{eff}^{(3)}=\tfrac{1}{9}J_d$.
Hence, in this case all the effective couplings along the bonds of the 
honeycomb lattice formed by the effective trimer spins are positive, and the ground state is a long-range ordered  antiferromagnet of effective trimer spins. 

\subsubsection{Equal trimer couplings}

\noindent In the case of equal intratrimer couplings, $J_1=J_2=J_3$, the lowest energy subspace consists of two degenerate SU(2) doublets. In addition to an effective spin-1/2 degree of freedom, an additional pseudospin degree of freedom  distinguishes these two doublets. Namely, the four ground states can be expressed~\cite{Subrahmanyam95,Wang2001} in the eigenbasis of the chirality operator
\begin{equation}\nonumber
\tau_{\Delta}^z \coloneqq -\frac{\sqrt{3}}{4}  \mathbf{S}_{1,\Delta} \cdot \left( \mathbf{S}_{2,\Delta} \times \mathbf{S}_{3,\Delta} \right)
\end{equation}
as
\begin{eqnarray}
\ket{+,\tfrac{1}{2}} &=& \frac{1}{\sqrt{3}} \left( \ket{\uparrow_1 \uparrow_2 \downarrow_3} + \omega\ket{\uparrow_1 \downarrow_2 \uparrow_3 }  + \omega^{*} \ket{\downarrow_1 \uparrow_2 \uparrow_3 }  \right), \nonumber\\
\ket{+,-\tfrac{1}{2}} &=& \frac{1}{\sqrt{3}} \left( \ket{\downarrow_1 \downarrow_2 \uparrow_3} + \omega\ket{\downarrow_1 \uparrow_2 \downarrow_3 }  + \omega^{*} \ket{\uparrow_1 \downarrow_2 \downarrow_3 }  \right), \nonumber\\
\ket{-,\tfrac{1}{2}} &=& \frac{1}{\sqrt{3}} \left( \ket{\uparrow_1 \uparrow_2 \downarrow_3} + \omega^{*}\ket{\uparrow_1 \downarrow_2 \uparrow_3 }  + \omega \ket{\downarrow_1 \uparrow_2 \uparrow_3 }  \right), \nonumber\\
\ket{-,-\tfrac{1}{2}}& =& \frac{1}{\sqrt{3}} \left( \ket{\downarrow_1 \downarrow_2 \uparrow_3} + \omega\ket{\downarrow_1 \uparrow_2 \downarrow_3 }  + \omega^{*} \ket{\uparrow_1 \downarrow_2 \downarrow_3 }  \right), \nonumber
\end{eqnarray}
with $\omega = e^{2\pi i /3}$, i.e., these states are chirality eigenstates  with  $\tau_{\Delta}^z \ket{\pm,m}=\pm \ket{\pm,m}$, for both $m=\pm\tfrac{1}{2}$, as well as spin-$S^z_\Delta$ eigenstates with $S^z_\Delta \ket{\pm,m}=m \ket{\pm,m}$. The same chirality operator and spin-chirality basis can be constructed for the right timers $\Delta'$.
From the $16\times 16$ matrix of the intertrimer interactions calculated in this basis, we  obtain the effective Hamiltonian 
\begin{align}
H_\mathrm{eff} =  \frac{J_{d}}{9} &\sum_{\Vec{r}}
A_{\Delta}^{(1)} A_{\Delta'}^{(1)} \mathbf{s}_{\Delta'}^{\Vec{r}} \cdot \mathbf{s}_{\Delta} ^{\Vec{r} + \Vec{a_1}}+A_{\Delta}^{(2)} A_{\Delta'}^{(2)} \mathbf{s}_{\Delta'}^{\Vec{r}} \cdot \mathbf{s}_{\Delta} ^{\Vec{r} + \Vec{a_2}} \nonumber \\
&\:\:\:\:\:\:+ A_{\Delta}^{(3)} A_{\Delta'}^{(3)} \mathbf{s}_{\Delta} ^{\Vec{r}}\cdot \mathbf{s}_{\Delta'}^{\Vec{r}} 
\label{eq:Heff_HomInteractions}
\end{align}
where the operators $A_{\Delta}^{(i)}$ act on the chirality degree of freedom on each trimer $\Delta$, and read 
\begin{align}
 A_{\Delta}^{(1)} &= 1 - 2\omega^{*}\tau_{\Delta}^{+} - 2\omega\tau_{\Delta}^{-}, \nonumber \\
A_{\Delta}^{(2)} &= 1 - 2\omega\tau_{\Delta}^{+} - 2\omega^{*}\tau_{\Delta}^{-}, \\
A_{\Delta}^{(3)} &= 1 - 2\tau_{\Delta}^{+} - 2\tau_{\Delta}^{-},\nonumber 
\end{align}
where $\tau^{+} = \ketbra{+}{-}$ and $\tau^{-} = (\tau^{+})^{\dagger}=\ketbra{-}{+}$, and likewise for $\Delta'$.
In the following, we do not consider further the chirality degree of freedom that emerges in the case of equal intratrimer couplings. It would certainly be interesting to address also possible chirality orders at this special high symmetry point in future work (however, an unbiased analysis of the effective model is unfeasible by QMC methods due to the sign problem~\cite{Weber2022A}).

\subsubsection{General trimer couplings}\label{Sec:EffHam_SpinHalf}

\begin{figure}[t]
    \centering
    \includegraphics[width=0.4\textwidth]{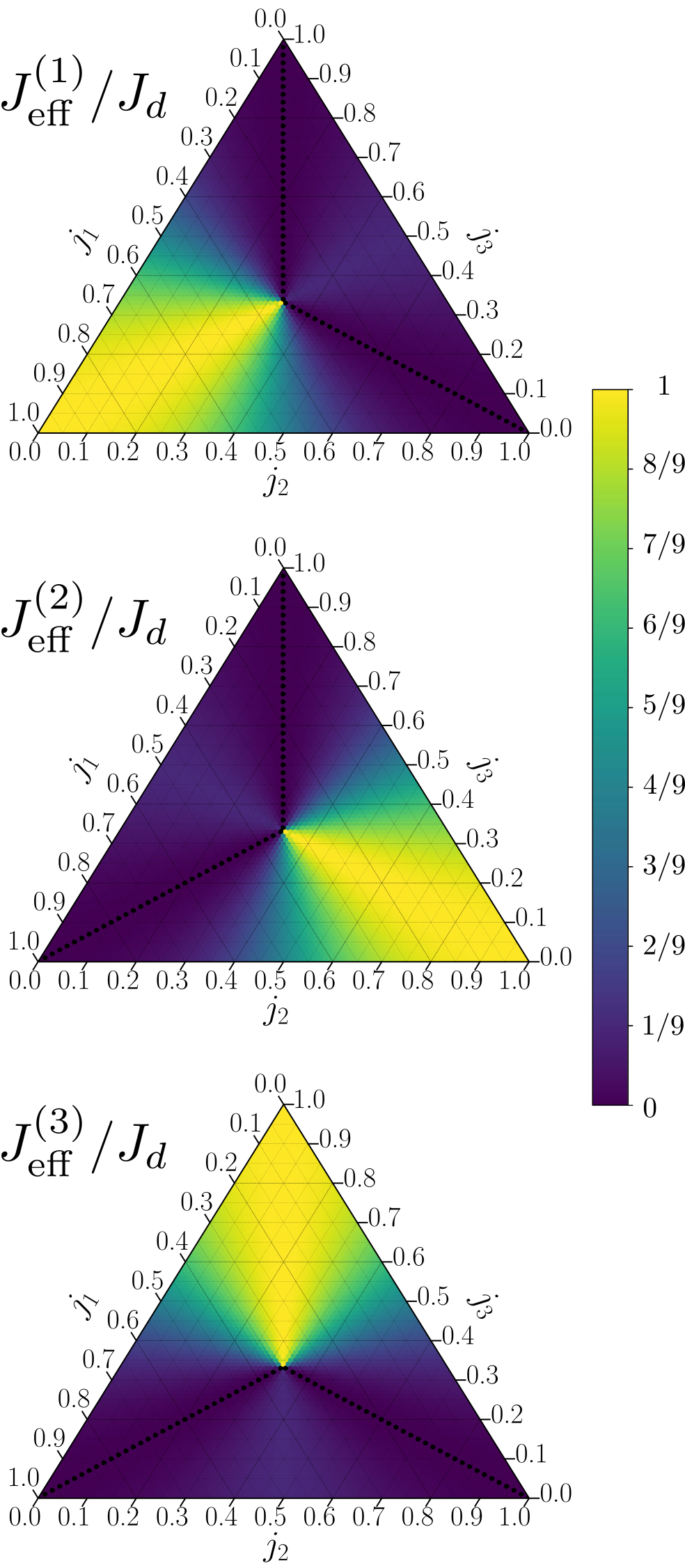}
    \caption{Dependence of the strength of the three effective Heisenberg couplings $J_\mathrm{eff}^{(1)}$, $J_\mathrm{eff}^{(2)}$ and $J_\mathrm{eff}^{(3)}$ on the intratrimer couplings  of the spin-1/2 Heisenberg model on the star lattice, as derived from perturbation theory in the weak intertrimer coupling  ($J_d$) limit.}
    \label{Fig:Jeffall}
\end{figure}

We can conveniently present the results from the perturbative calculation of the effective coupling strengths between the trimer spins in the parameter space of the intertrimer couplings $J_1$, $J_2$, and $J_3$ using barycentric coordinates. These are defined as $j_i=J_i/(J_1+J_2+J_3)$, such that $j_1+j_2+j_3=1$. In the following, we denote the sum of the three intertrimer couplings by $J_\Delta=J_1+J_2+J_3$, such that $j_i=J_i/J_\Delta$. 
The values of the three effective couplings $J_\mathrm{eff}^{(i)}$ are shown in Fig.~\ref{Fig:Jeffall} using the above defined barycentric coordinates. 

We observe that over large regions of parameter space a single effective coupling dominates. For example, for strong $J_3$, the effective coupling $J_\mathrm{eff}^{(3)}$ dominates, and similarly for strong $J_1$ ($J_2$), it is $J_\mathrm{eff}^{(1)}$ ($J_\mathrm{eff}^{(2)}$) that is the largest. In terms of the effective total trimer spins, which reside on the effective honeycomb lattice, this leads to a strong tendency to form a quantum disordered ground state in the presence of a weak $J_d$. 
Indeed, if one of the three effective couplings dominates, there is a strong tendency to form singlets between pairs of trimer spins that are connected by those strongest effective couplings. Since each trimer spin is connected to a single other trimer spin by one of these couplings, the original spin system becomes effectively dimerized on the level of the effective trimer spins.

We will further detail the resulting phase diagram of the spin-1/2 Heisenberg model on the star lattice in the weak-$J_d$ regime after we compare, in the following section, the thermodynamic predictions from the above effective Hamiltonians to the QMC data in the weak intertrimer coupling regime, based on the trimer basis.

\subsection{QMC results for weak dimer coupling}

In order to compare the effective Hamiltonian description to the thermodynamics of the underlying star lattice model in the weak-$J_d$ regime, we performed QMC simulations for different parameter values, fixing  $J_d=0.05 J_\Delta$. This value of $J_d$ indeed corresponds to the ratio of the dimer and trimer couplings estimated for the compound 
[(CH${}_3$)${}_2$(NH${}_2$)]${}_3$[Cu${}_3$($\mu_3$-OH)($\mu_3$-SO${}_4$)($\mu_3$-SO${}_4$)${}_3$]·0.24H${}_2$O in Ref.~\cite{Sorolla2020}.
Defining $j_d=J_d/J_\Delta$, this means that we consider $j_d=0.05$ in the following.
\begin{figure}[t]
    \centering

    \vspace{-0.7cm}
    
    \includegraphics[width=0.5\textwidth]{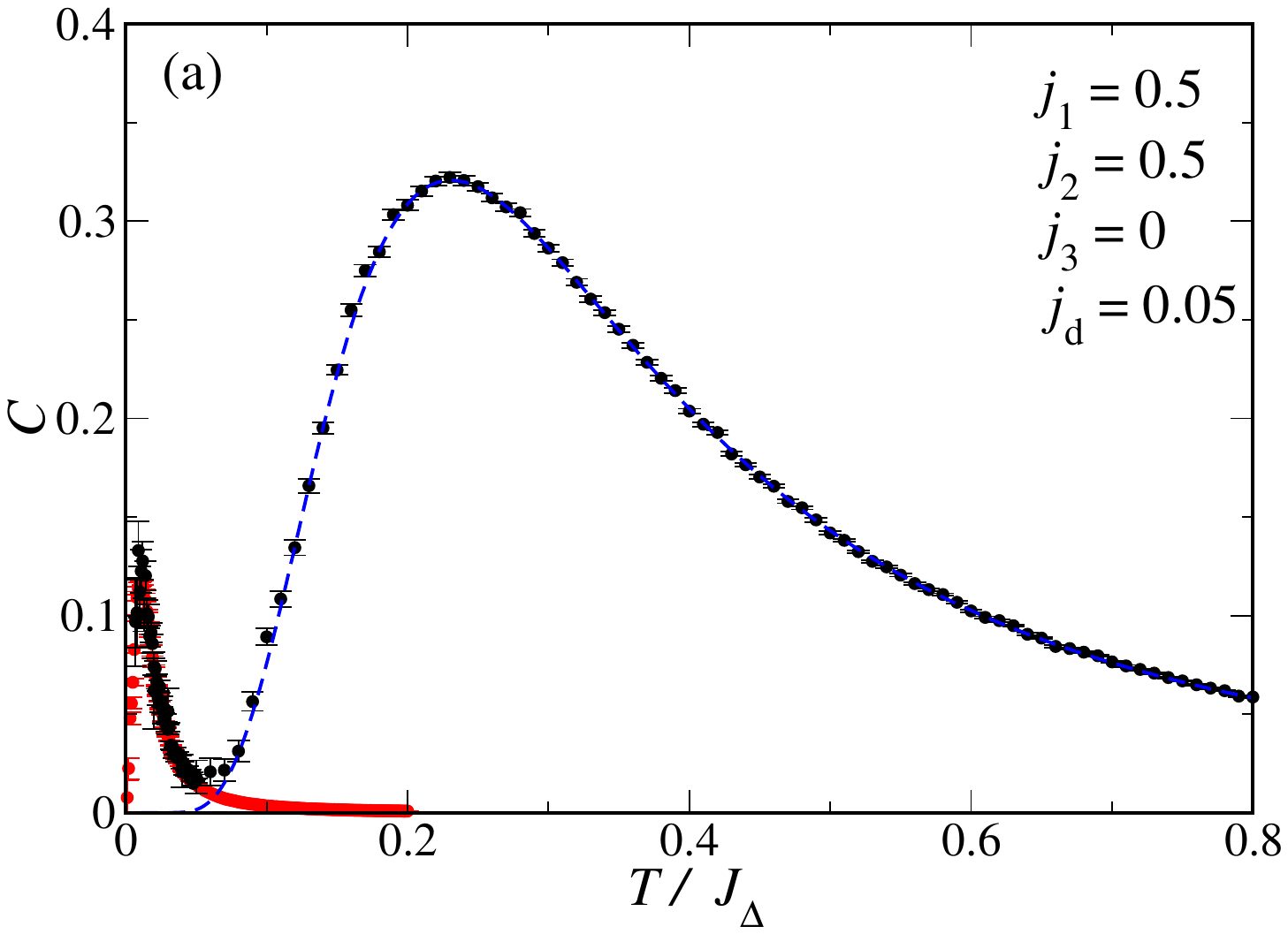}

    \vspace{-1cm}
    
    \includegraphics[width=0.5\textwidth]{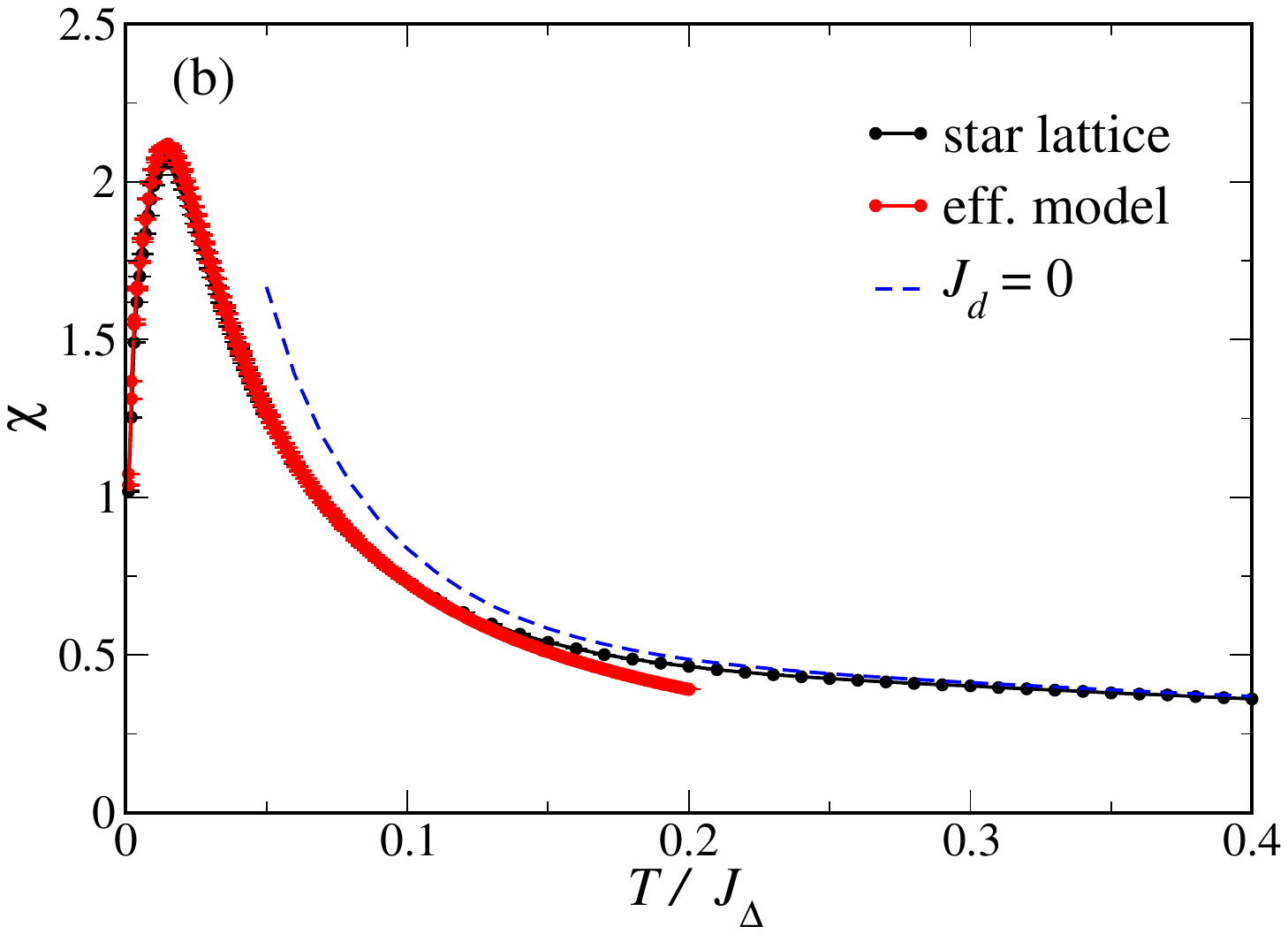}
    \caption{Temperature dependence of (a) the specific heat $C$ and (b) the uniform susceptibility $\chi$ of the spin-1/2 Heisenberg model on the star lattice (black symbols) and for the effective low-energy spin-1/2 Heisenberg model on the honeycomb lattice (red symbols) for the parameter choice (i),  with $L=16$. Dashed blue lines show the results for isolated trimers ($J_d=0$), obtained from ED.}
    
    \label{Fig:compare1}
\end{figure}

\begin{figure}[t]
    \centering

    \vspace{-0.7cm}
    
    \includegraphics[width=0.5\textwidth]{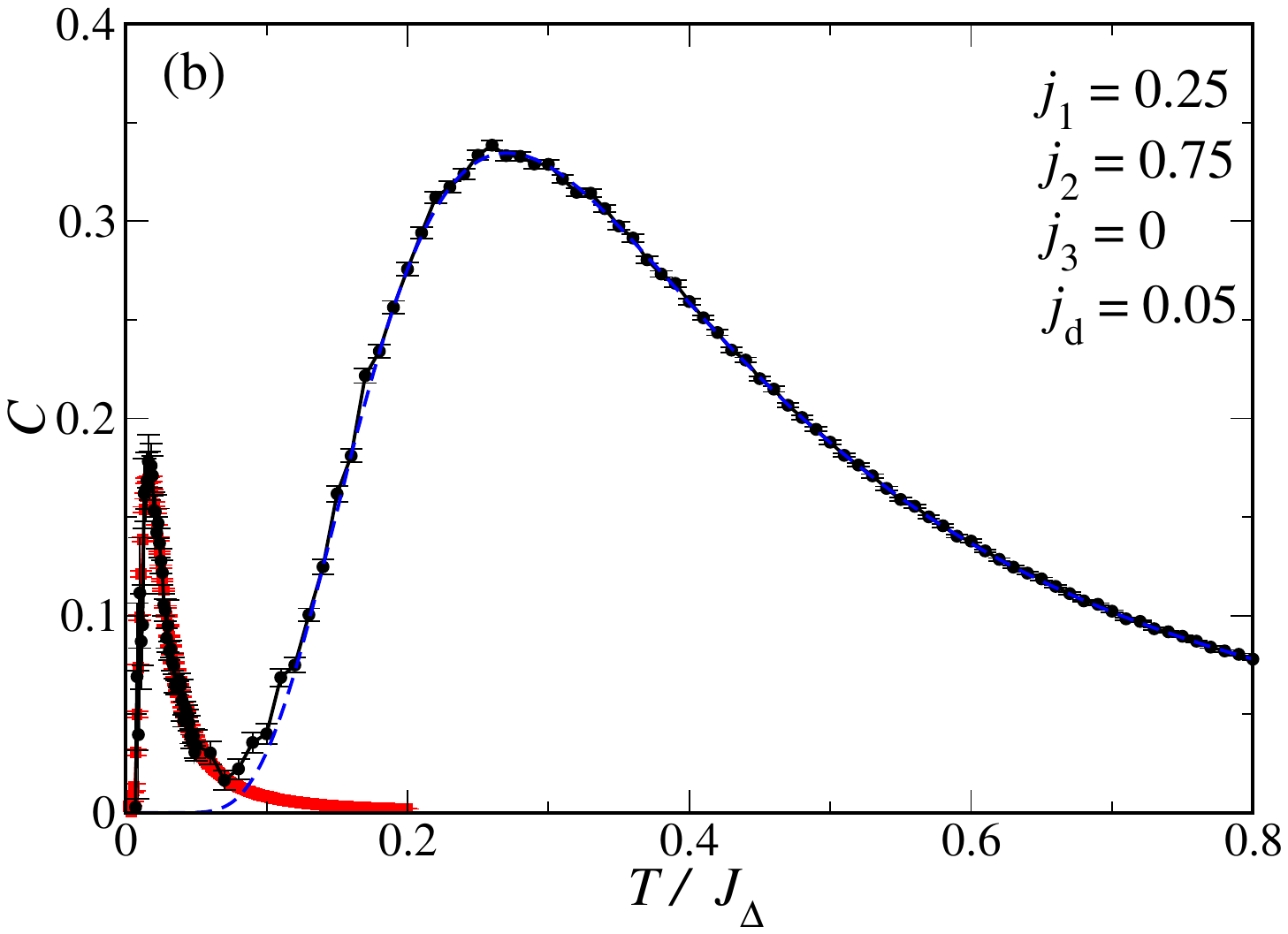}

    \vspace{-1cm}
    
    \includegraphics[width=0.5\textwidth]{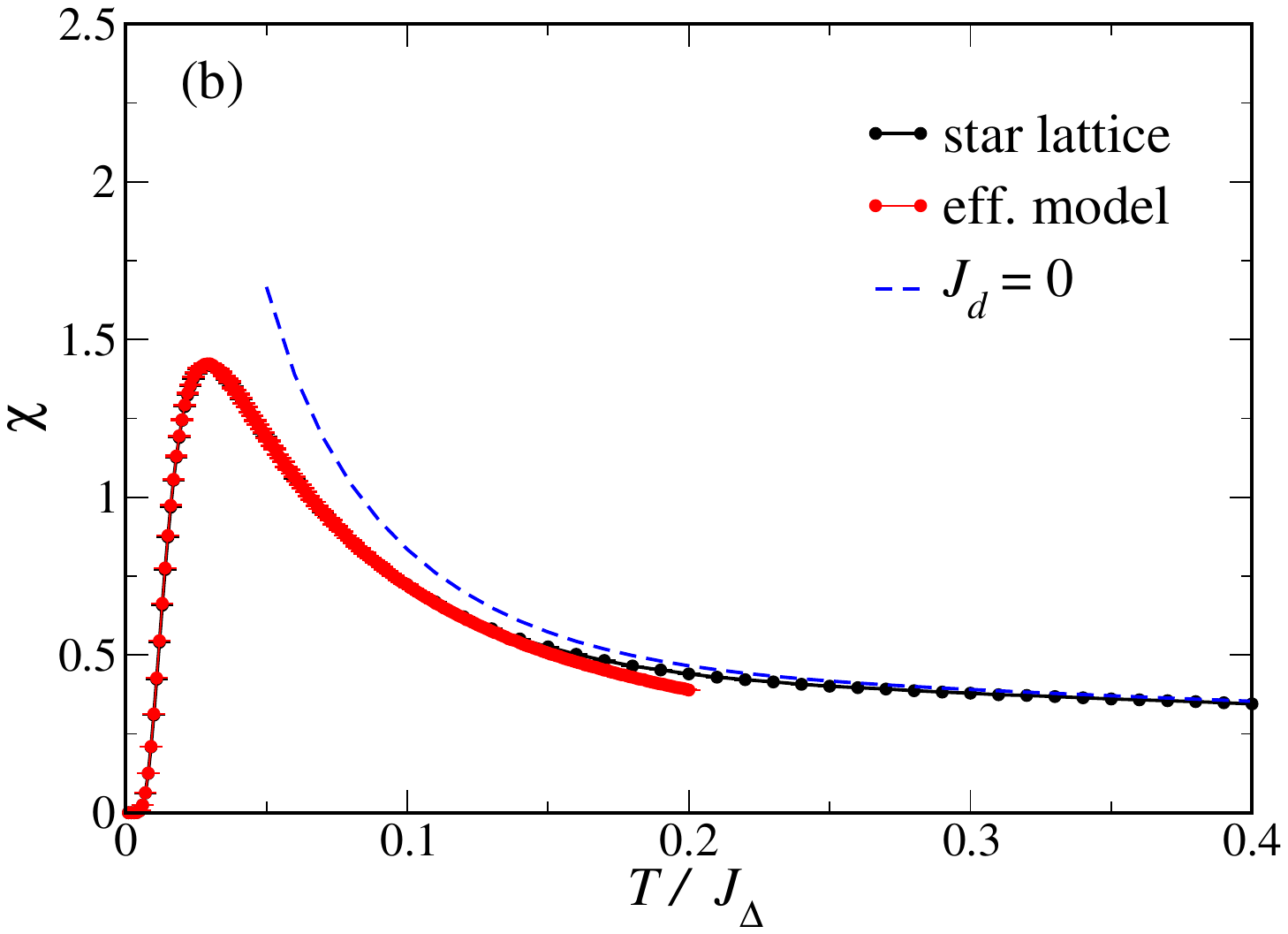}
    \caption{Temperature dependence of (a) the specific heat $C$ and (b) the uniform susceptibility $\chi$ of the spin-1/2 Heisenberg model on the star lattice (black symbols) and for the effective low-energy spin-1/2 Heisenberg model on the honeycomb lattice (red symbols) for the parameter choice (ii), with $L=16$. Dashed blue lines show the results for isolated trimers ($J_d=0$), obtained from ED.}
    \label{Fig:compare2}
\end{figure}

A quantitative comparison of the thermodynamic properties of the star lattice model with the effective honeycomb lattice model requires to probe the thermodynamic properties of the star lattice model on temperature scales of the order and below the strength of the dimer coupling $J_d$, since the effective couplings on the honeycomb lattice scale with $J_d$. QMC simulations at such low temperatures are not feasible in those regime where the simulations are plagued by the sign problem. In order to test the accuracy of the effective model for the low temperature properties, we thus consider the star lattice model in specific parameter regimes, where the sign problem can be completely avoided. This is granted in particular along the outer borders of the triangles of the barycentric plots in Fig.~\ref{Fig:Jeffall}, corresponding to the cases where one of the trimer couplings $J_i$ vanishes. Indeed, the star lattice model becomes bipartite in these limits. More specifically, we consider the following two  cases: (i) $j_1=j_2=0.5$, $j_3=0$, and (ii) $j_1=0.25$, $j_2=0.75$, $j_3=0$. With the above choice of $j_d=0.05$, we obtain for the effective couplings: (i)
$j_\mathrm{eff}^{(1)}=j_\mathrm{eff}^{(1)}=0.0\bar{2}$, $j_\mathrm{eff}^{(3)}=0.00\bar{5}$ and (ii)
$j_\mathrm{eff}^{(1)}=0.0463949$,
$j_\mathrm{eff}^{(2)}=0.0021496$, and
$j_\mathrm{eff}^{(3)}=0.0014555$, respectively, where we defined $j_\mathrm{eff}^{(i)}=J_\mathrm{eff}^{(i)}/J_\Delta$. We performed QMC simulations for both the original star lattice model as well as the effective spin-1/2 model on the honeycomb lattice in order to compare their thermodynamic properties. Such comparisons are shown for both parameter cases for the uniform susceptibility $\chi$ and the specific heat $C$ in Fig.~\ref{Fig:compare1} (case (i)) and Fig.~\ref{Fig:compare2} (case (ii)). Here, both intensive quantities have been normalized per site of the original star lattice model, and the simulations were performed on $L=16$ lattices. We find that the low-temperature behavior of the star lattice model is rather well described by the effective model - in particular, the latter captures well the low-$T$ behavior in both quantities. The specific heat of the star lattice model exhibits in both cases a two-peak structure:  The low-$T$ peak relates to the onset of intertrimer correlations, which are captured by the effective honeycomb model, while the second peak stems from the release of entropy upon increasing $T$ by suppressing intratrimer spin correlations, as seem from comparing to the corresponding results for $C$ for $J_d=0$, which are included in both figures. Therefore, this second peak falls beyond the scope of the effective model and instead relates to the intratrimer physics. From the low-$T$ behavior of the susceptibility we can furthermore infer ground state properties: In the parameter case (i), the susceptibility approaches towards a finite ground state value, indicative of antiferromagnetic order, while in the second case (ii), the susceptibility is strongly suppressed towards zero temperature, indicating a finite magnetic excitation gap such as in a quantum disordered ground state. 
These findings are indeed in accord with the overall phase diagram that we extract from the effective honeycomb lattice model, as detailed in the following section. 

\subsection{Phase diagram for weak dimer coupling} 

\begin{figure}[t]
    \centering
    \includegraphics[width=0.5\textwidth]{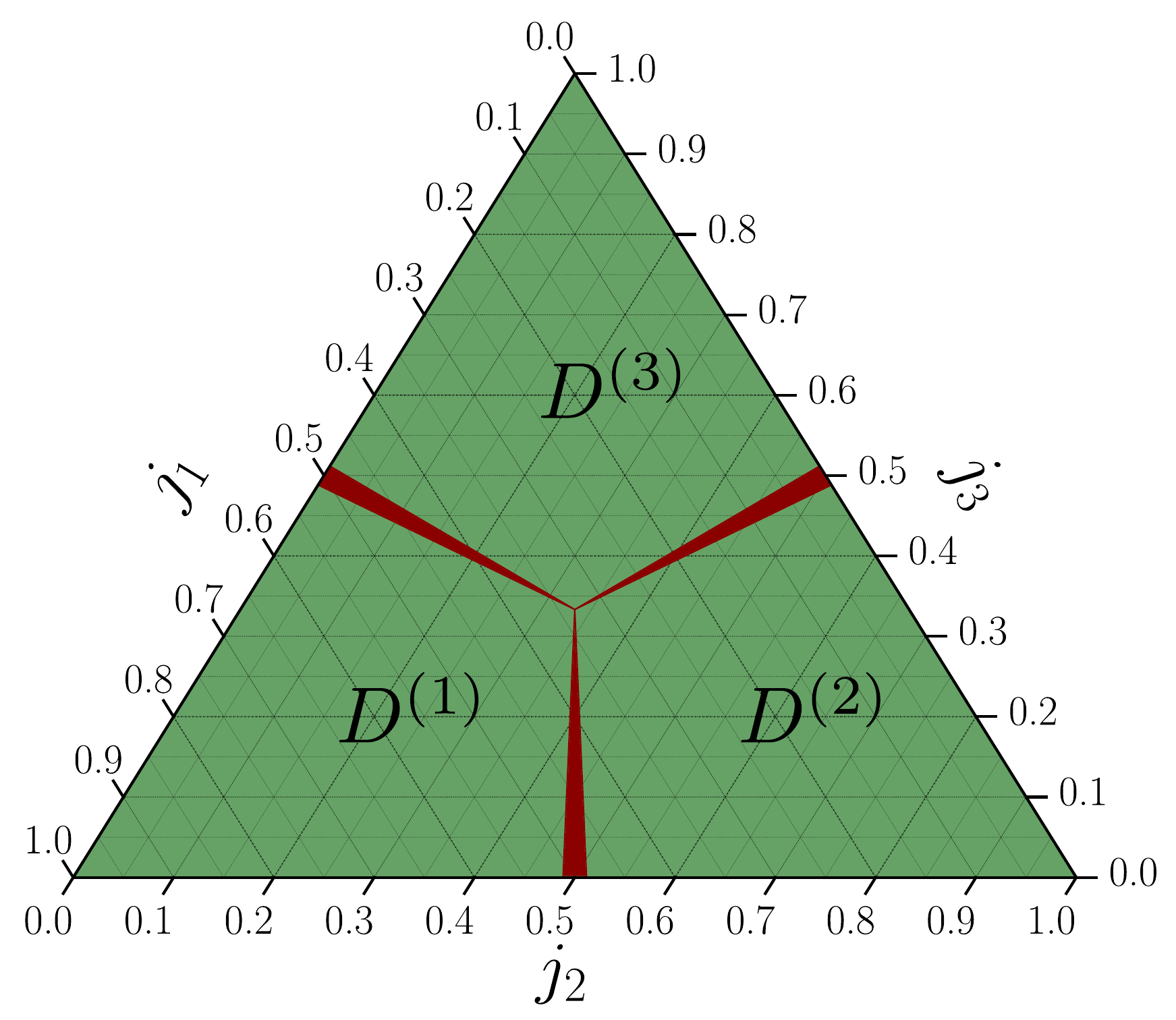}
    \caption{Phase diagram of the star lattice model in the weak $J_d$ limit, in dependence on the trimer couplings. Within the narrow red regions, the effective trimer spins form a long-range antiferromagnetically ordered ground state. Beyond the antiferromagnetic domain (red), the predominance of a single effective coupling in each unit cell leads to the formation of three distinct quantum disordered regions (green) with dominant singlets forming along the intertrimer bonds with the strongest effective coupling, indicated by $D^{(1)}$, $D^{(2)}$ and $D^{(3)}$, respectively.}
    \label{Fig:phase_effHoney}
\end{figure}

Based on the effective honeycomb models derived above, we can construct the phase diagram of the spin-1/2 Heisenberg model on the star lattice within the weak dimer coupling regime. To do so, we use recent results for the ground state phase diagram of the spin-1/2 Heisenberg model on the honeycomb lattice with three different coupling strengths along the three bond directions~\cite{Sushchyev2023}. This allows us to extract the nature of the ground state in the weak-$J_d$ region based on the values of the effective couplings  $J_\mathrm{eff}^{(i)}$, expressed in terms of $j_1,j_2$ and $j_3$. The resulting phase diagram is shown in Fig.~\ref{Fig:phase_effHoney}, where we identify four distinct regions: For most parameter values, the predominance of one of the three effective couplings $J_\mathrm{eff}^{(i)}$, $i=1,2$ or $3$, leads to the formation of a quantum disordered ground state, denoted by $D^{(i)}$, with a strong formation of singlets among neighboring trimer spins in the respective lattice direction. Only within a rather narrow range of parameter values where at least two of the trimer couplings are of similar strength there exists a ground state with long-range antiferromagnetic ordering of the trimer spins. 

\begin{figure}
    \centering

    \vspace{-0.7cm}
    
    \includegraphics[width=0.5\textwidth]{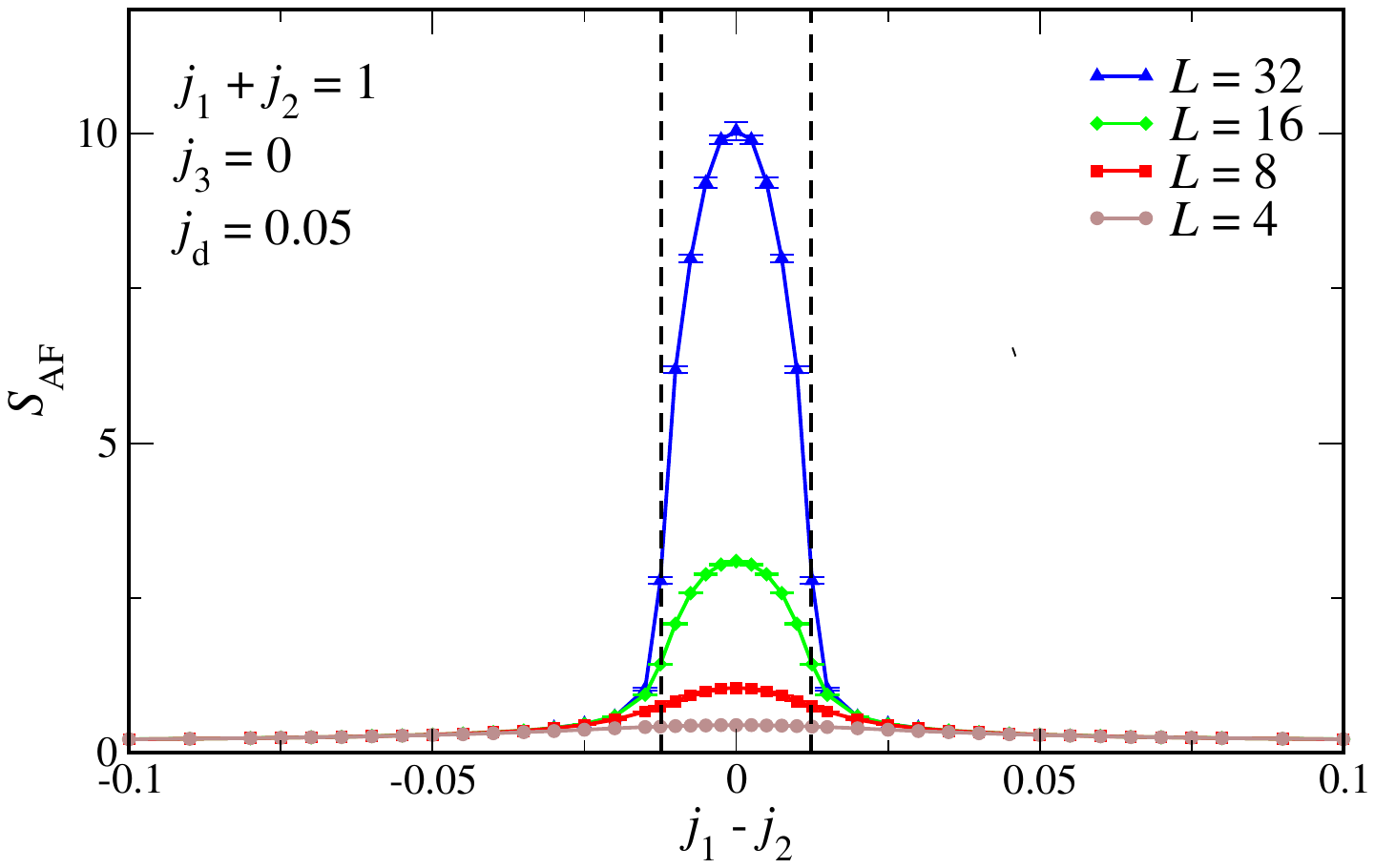}
    \caption{Scan of the staggered structure factor $S_\mathrm{AF}$ of the trimer spins of the star lattice model along the parameter line with $j_3=0$, and $j_d=0.05$ as a function of the difference $j_1-j_2$. The two vertical lines indicate the borders of the antiferromagnetic regime as derived from the effective honeycomb lattice model.}
    \label{Fig:phasecut}
\end{figure}

We can  compare the predicted ground state phase diagram with QMC results by concentrating again on the edges of the phase diagram, along which one of the couplings $J_i$ vanishes, leading to a bipartite model and the absence of a sign problem. More specifically, we focus on the line $j_3=0$, such that $j_1+j_2=1$. In Fig.~\ref{Fig:phasecut} the value of the low-temperature antiferromagnetic structure factor $S_\mathrm{AF}$ of the trimer spins, as measured for star lattices of increasing size $L$, is shown as a function of the difference $j_1-j_2$. We find that only within a rather restricted regime around the point $j_1=j_2$ does the structure factor grow with increasing system size, indicative of antiferromagnetic long-range order in the ground state, while it becomes rapidly suppressed at finite differences between $j_1$ and $j_2$ beyond about $0.0125$. The dashed vertical lines in  Fig.~\ref{Fig:phasecut} display the border lines of the antiferromagnetic regime taken from the overall phase diagram (Fig.~\ref{Fig:phase_effHoney}), and they again demonstrate a remarkably good agreement between the star lattice model QMC results and the effective honeycomb lattice model for the low energy properties.

Finally, we shall relate the above phase diagram to previous iPEPS results for the isotropic case, $J_1=J_2=J_3$~\cite{Jahromi2018} in the weak-$J_d$ region: The VBS ground state manifold identified in Ref.~\cite{Jahromi2018} is characterized by a six-site unit cell and  a predominant singlet forming on a single bond within each trimer. In Ref.~\cite{Jahromi2018}, this VBS ground state manifold is reported to be three-fold degenerate. However, a closer inspection of the three degenerate VBS states illustrated in Fig.~12 of Ref.~\cite{Jahromi2018} shows that this count is incomplete. For example, the VBS state shown in Fig.~12a of Ref.~\cite{Jahromi2018} is characterized by the predominant formation of singlets along one of the three parallel trimer bond directions, e.g., those corresponding to $J_1$ in our notation. Apparently then there must be two further, equivalent such states, with predominant singlet formation on the trimer bonds corresponding to $J_2$ and $J_3$, respectively. Similarly, the patterns illustrated in Fig.~12b and c of Ref.~\cite{Jahromi2018} can be locally rotated to give rise to four additional states and hence in total the six-site unit cell VBS ground state is nine-fold degenerate: In terms of dominant singlets, each of these nine states corresponds to picking one trimer bond from each of the two trimers $\Delta$ and $\Delta'$ within each unit cell. Such a ground state manifold can be perturbed upon breaking explicitly the symmetry among  the trimer bonds. In particular, upon enhancing the strength of, e.g., all $J_1$ trimer bonds leads to the preference within the VBS ground state manifold  of the single specific state wherein the predominant singlets reside on the $J_1$-bonds, and similarly for $J_2$ and $J_3$. Thus, we arrive at the following simple picture: Starting from the VBS ground state for the isotropic limit, the presence of a single dominant trimer coupling $J_i$ leads into the corresponding regime $D^{}(i)$, in accord with the above phase diagram. Certainly, a similar stabilization of any other remaining six VBS ground state would require to break up the symmetry among the trimer bonds even further (i.e., by picking a strong bond $J_i$ on the left trimer and a strong bond $J_j$ with $i\neq j$ on the right trimer in each unit cell). Such explicitly symmetry broken states however reside beyond the inversion symmetric case considered here and therefore are not contained within the above phase diagram.  

\section{Conclusions}\label{Sec:Conclusions}
In summary, we used a combination of ED, QMC, FTLM and effective Hamiltonian approaches to study the thermodynamic properties of the spin-1/2 Heisenberg model on the star lattice. Besides the Heisenberg model on the kagome lattice, this system is the only other SU(2)-symmetric Archimedean lattice model that features a non-magnetic, quantum disordered ground state. 
Due to the strong geometric frustration in this system, QMC simulations suffer from a severe sign problem, restricting efficient QMC calculations to the regime of strong dimer coupling (using the dimer or local site basis) or weak dimer coupling (using the trimer basis). 

However, from our combined analysis we were able to obtain the thermodynamic properties in the regime of dominant dimer coupling, including the special point of balanced dimer and trimer couplings ($J_d=J_t$), where the system is characterized by a non-magnetic, quantum disordered ground state with a strong dimerization on the dimer bonds and a sizeable spin gap (singlet-triplet gap). This leads to a pronounced activated behavior in the thermodynamic response at low temperatures, and in contrast to other strongly frustrated quantum spin models no low-energy singlet excitations reside below the spin gap.  

In the regime of weak dimer coupling, we derived effective low-energy Hamiltonians on the honeycomb lattice of trimers by considering the general case with three different trimer couplings $J_1$, $J_2$, and $J_3$ and inversion symmetry. From explicit QMC simulations, we found that the low-temperature thermodynamics of the weakly coupled trimer system is indeed well described by these effective Hamiltonians. The ground state phase diagram in the regime of weak dimer coupling is  dominated by three extended quantum disordered regimes, each being characterized by  the dimerization of the effective trimer spins along a specific lattice direction of the honeycomb lattice. Only within three rather narrow regimes, in which two of the trimer couplings come close, such as for $J_1\approx J_2$, does the system exhibit an ordered ground state of antiferromagnetically aligned trimer spins. The VBS state identified from iPEPS calculations in Ref.~\cite{Jahromi2018} for the isotropic case, $J_1=J_2=J_3$, also fits well into this picture, once its full degeneracy is accounted for. It would nevertheless be interesting to examine in future studies, whether the VBS ground state is obtained also from the effective Hamiltonian, including the additional chirality degree of freedom, that we derived for the case of equal trimer couplings. 

As  mentioned in the introduction, recently a  Cu-based  spin-1/2 quantum spin system with an underlying star lattice geometry was reported for the layered compound
[(CH${}_3$)${}_2$(NH${}_2$)]${}_3$[Cu${}_3$($\mu_3$-OH)($\mu_3$-SO${}_4$)($\mu_3$-SO${}_4$)${}_3$]·0.24H${}_2$O in Ref.~\cite{Sorolla2020}. Based on ab-initio calculations, the ratio of the trimer and dimer coupling strength was estimated to be of order 20, such that the quantum spin system resides well in the regime of weak dimer couplings. From our analysis, combined with the iPEPS results of Ref.~\cite{Jahromi2018}, the reported absence of magnetic order in this compound can indeed be expected in this parameter regime. Furthermore, our results suggest that upon applying uniaxial pressure, it might be feasible to drive the material across the phase boundary of the antiferromagnetic regions, in case the trimer couplings are sufficiently affected by the corresponding lattice distortions. Certainly, further ab-initio analysis would be required in order to  explore this possibility for this specific compound.

\section*{Acknowledgements}
We thank Andreas Honecker and Alexander Sushchyev for useful discussions. Furthermore, we acknowledge support by the Deutsche Forschungsgemeinschaft (DFG) through the RTG 1995, and thank the IT Center at RWTH Aachen University for access to computing time through the JARA Center for Simulation and Data Science. A. R. also thanks the DFG funded SFB 1170 on Topological and Correlated Electronics at Surfaces and Interfaces.

\bibliography{main.bbl}

\end{document}